\documentclass[%
 reprint,
 superscriptaddress,
 longbibliography,
 amsmath,amssymb,
 aps, pra,
]{revtex4-2}

\usepackage{graphicx}%
\usepackage{dcolumn}%
\usepackage{bm}%
\usepackage{bbm}%
\usepackage{tabularx}
\usepackage{array}
\usepackage{orcidlink}

\usepackage{hyperref}%
\hypersetup{
	colorlinks=true,
	linkcolor=blue,
	citecolor=blue,
	urlcolor=blue,
}

\usepackage[dvipsnames]{xcolor}

\usepackage{mathtools}

\newcommand{\fc}[1]{{#1}}

\newcommand{\nparam}{n_\text{param}}
\newcommand{\ngate}{n_\text{param}}

\begin{document}

\title{Backpropagating Pauli Propagation}

\newcommand{\QuantinuumGermany}{\affiliation{Quantinuum, Leopoldstra{\ss}e 180, 80804 Munich, Germany.}}

\author{Sheng-Hsuan~Lin\,\orcidlink{0000-0002-8599-0378}}\email{sheng-hsuan.lin@quantinuum.com} \QuantinuumGermany
\author{Etienne~Granet\,\orcidlink{0000-0002-6594-7022}}\QuantinuumGermany
\author{Kevin~H\'emery\,\orcidlink{0000-0001-7086-593X}}\QuantinuumGermany
\author{Henrik~Dreyer\,\orcidlink{0000-0002-1480-6406}}\QuantinuumGermany

\begin{abstract}
We develop a backpropagation algorithm for evaluating parameter gradients in quantum circuits using Pauli propagation simulation.
The method has computational complexity comparable to that of standard sparse Pauli simulation techniques, while producing gradients whose accuracy is of the same order as the corresponding observable expectation values.
By exploiting the reversibility of quantum circuits, the algorithm  reduces the memory cost by a factor of $\mathcal{O}(n_\text{param})$ compared with conventional reverse-mode automatic differentiation, where $n_\text{param}$ denotes the number of parameters in the circuit. 
Compared with finite difference methods, the algorithm is $\mathcal{O}(n_\text{param})$ more efficient in function evaluations. 
These features enable efficient and accurate classical optimization of quantum circuits for applications such as state preparation and time-evolution compression, while also allowing operator-complexity measures such as the operator stabilizer R\'enyi entropy to be monitored and regularized during optimization.
We demonstrate the method by optimizing low-energy state-preparation circuits for transverse-field Ising models in one, two, and three dimensions and for the three-dimensional Heisenberg model, and by compressing two-dimensional time-evolution circuits.

\end{abstract}

\maketitle

\section{\label{sec:intro} Introduction}

Quantum computers have the potential to solve problems that are intractable for classical devices, with quantum simulation widely regarded as one of their most promising applications. A number of quantum algorithms can be implemented without any classical optimization. For example, the ground state energies can be estimated by combining adiabatic state preparation with quantum phase estimation. Nevertheless, incorporating classical optimization during circuit design can often significantly improve algorithmic performance and reduce the quantum resources required. Important examples include variational state preparation and variational compression of time-evolution circuits.

Even though the simulation of general quantum systems is exponentially hard, many physically relevant cases admit efficient classical descriptions, typically when the available quantum resources --- such as entanglement, non-stabilizerness, or non-Gaussianity --- are limited.
In the Schr\"odinger picture, this allows efficient simulation of quantum states: tensor network state methods~\cite{bridgeman2017hand,cirac2021matrix} exploit low entanglement, stabilizer-based methods~\cite{aaronson2004improved,bravyi2019simulation} exploit limited non-stabilizerness, and free-fermion techniques~\cite{surace2022fermionic} exploit limited non-Gaussianity.
Each of these classical methods has been used for circuit optimization with different strengths and limitations.
In the Heisenberg picture, simulations are based on tracking the time evolved operator $O(t) = U^\dagger(t) O U(t)$ and expectation values of observables can be evaluated at the end of the evolution.
Tensor network operator methods~\cite{anand2023classical,beguvsic2024fast} leverage low operator entanglement~\cite{prosen2007operator}, sparse Pauli dynamics methods~\cite{beguvsic2025simulating} exploit limited operator non-stabilizerness~\cite{dowling2025magic}, and sparse Majorana methods~\cite{miller2025simulation} exploit limited operator non-Gaussianity~\cite{debertolis2025natural}.
In stark contrast to the Schr\"odinger picture, where these resources typically lead to distinct simulable regimes, recent work has shown that the corresponding non-classical resources in the Heisenberg picture are quantitatively connected~\cite{dowling2025bridging}.
This motivates Heisenberg-picture simulation as a natural framework for observable estimation and, potentially, for circuit optimization.

\fc{Pauli propagation originates from early Fourier-based analyses showing that noise suppresses high-complexity components of quantum circuits~\cite{gao2018efficient}, through stochastic Pauli estimators for observable expectation values~\cite{rall2019simulation}, into a broader framework for efficiently approximating quantum dynamics in the Pauli basis~\cite{rudolph2025pauli}. Beyond its use as a theoretical tool for establishing classical-simulability results~\cite{aharonov2023polynomial,nemkov2023fourier,shao2024simulating,schuster2025polynomial,angrisani2025classically,fontana2025classical,cirstoiu2024fourier}, it has become an increasingly useful numerical method for quantum dynamics due to its conceptual simplicity, ease of implementation, and flexibility with respect to system geometry.
In this work, we focus on a deterministic sparse variant in which observables are propagated in the Heisenberg picture while their Pauli expansions are compressed according to a prescribed truncation rule, such as coefficient magnitude, Pauli weight, or a fixed sparsity budget; throughout, we refer to this approach as sparse Pauli dynamics (SPD)~\cite{beguvsic2024fast,beguvsic2025real}.}

Despite these advantages, the application of SPD to circuit optimization remains limited compared with tensor network operator approaches.
A key obstacle is the lack of an efficient and systematic method for computing gradients within the SPD framework, and existing optimization strategies have largely been ad hoc.
For example, some works rely on stochastic gradient approximations such as simultaneous perturbation stochastic approximation (SPSA) combined with classical optimizers~\cite{lin2026utility}, while others employ automatic differentiation through the full simulation pipeline~\cite{d2025circuit}, which requires storing all intermediate operators and can therefore incur significant memory overhead. Surrogate-based or symbolic approaches~\cite{fontana2025classical,rudolph2023classical,lerch2026efficient,monaco2025symbolic,li2026dual} have also been explored but are limited by the precompiled path.
These existing methods can provide approximate gradients, however, their computational and memory costs can become prohibitive for large circuits, limiting the practical use of SPD as a circuit-optimization tool.

\begin{figure*}[th!]
    \centering
    \includegraphics[width=\linewidth]{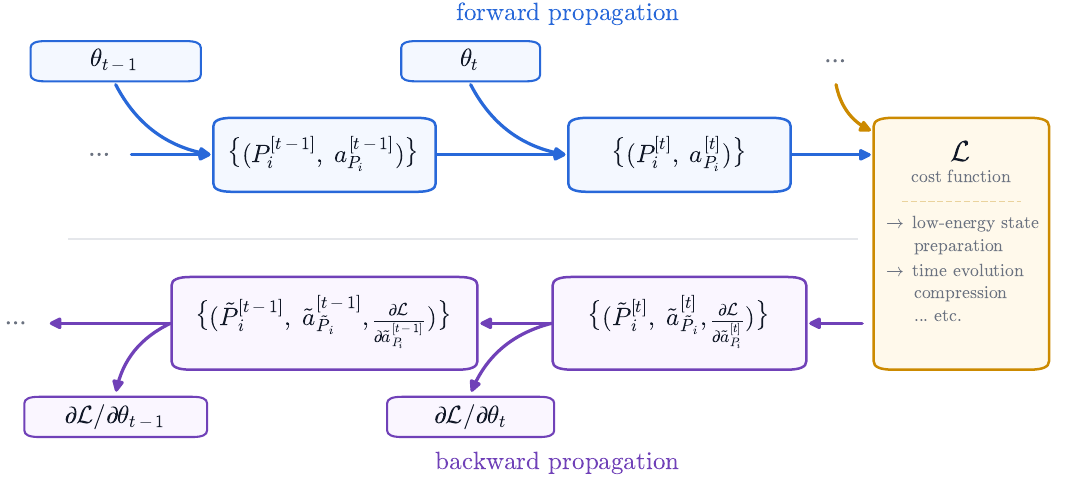}
    \caption{An illustration of the forward propagation and backward propagation workflow of the SPD optimization.
    The workflow resembles reverse-mode automatic differentiation on a computational graph. The difference is that we do not store values of the intermediate nodes. Instead, we utilize the reversibility of the quantum circuit and recompute the values on the nodes on the fly in the backward pass.
Our directional terminology follows the operator propagation rather than the Schr\"odinger-picture state evolution: ``forward'' means applying the Heisenberg updates $O_s=U_s^\dagger O_{s-1}U_s$ in circuit order, thereby growing the observable toward the circuit input, whereas ``backward'' means traversing the gates in reverse order and applying the inverse updates $O_{s-1}=U_sO_sU_s^\dagger$.
    }
    \label{fig: illustration}
\end{figure*}

In this work, we introduce a memory-efficient method for computing approximate gradients within the SPD framework.
Our main contribution is to show that reverse-mode gradient evaluation can be implemented without storing the full forward trajectory of intermediate operators. By exploiting the reversibility of quantum circuits, the required intermediate operators are recomputed during an additional backward pass. The proposed algorithm augments standard Pauli propagation with a backward propagation step, analogous in spirit to backpropagation in artificial neural networks. An illustration of the scheme is shown in Fig.~\ref{fig: illustration}.
Compared with standard finite-difference gradient methods, which require $\mathcal{O}(\nparam)$ evaluations, the backpropagation method hence has a $\mathcal{O}(\nparam)$ runtime speedup, where $\nparam$ is the number of parameterized gates.
Compared with conventional reverse-mode automatic differentiation, this reduces the memory cost by a factor of $\mathcal{O}(\ngate)$ while preserving the same asymptotic runtime and memory scaling as the original SPD simulation.
The main tradeoff is an additional truncation error arising from the recomputation of intermediate operators, but the resulting gradient accuracy remains comparable to the accuracy of the SPD estimate of the observable expectation value.
These features make the method particularly suitable for large-scale simulations where memory is the primary bottleneck.

The remainder of this paper is organized as follows. In Sec.~\ref{sec:methods}, we introduce the sparse Pauli decomposition framework, present our gradient evaluation method based on backward Pauli propagation, and compare it with other differentiation schemes.
In Sec.~\ref{sec:results}, we demonstrate the performance of our method on state preparation for the transverse-field Ising model in one, two, and three dimensions, the three-dimensional Heisenberg model, and on time-evolution circuit compression.
\fc{Throughout, we use 1D, 2D, and 3D to denote one-, two-, and three-dimensional systems, respectively.}
In Sec.~\ref{sec: discussion}, we discuss the computational complexity and implementation aspects, including regularization and parallelization. We conclude in Sec.~\ref{sec: conclusion}. Additional technical details, numerical data, and implementation notes are provided in the appendices.

\section{\label{sec:methods} Methods}

\subsection{\label{subsec: spd} Sparse Pauli Decomposition}

The Pauli operators form an orthonormal basis for the space of operators.
In general, an operator can be written as
\begin{equation}\label{eq: decomposition}
    O = \sum_{P \in \mathcal{P}} a_P P ,
\end{equation}
that is a linear combination of Pauli operators $P$ with complex coefficients $a_P$.
When $\mathcal{P}$ is the complete basis, this representation requires storing all $4^N$ coefficients $a_P$.
For a Hermitian operator, the coefficients are real numbers.

Often, when the operator is simple or has certain structure, the majority of the coefficients are zero.
The idea of SPD is to store only the contributing pairs $\{(P, a_P)\}$, which can significantly reduce the memory usage when the representation is actually ``sparse''.
This idea admits the same expression as in Eq.~\ref{eq: decomposition} with the difference that the set $\mathcal{P}$ is the set of Pauli operators that contribute to $O$.

In practice, an approximation with a finite truncation threshold $\delta$ is introduced. Namely, the sum includes only the term where $|a_P|>\delta$,
\begin{equation}\label{eq: sparse_pauli_decomposition}
    O \approx \sum_{P \in \mathcal{P}_\delta} a_P P ,
\end{equation}
where $\mathcal{P}_\delta = \{P\ \big|\ |a_P|>\delta \}$.
Below, we refer to such representation of an operator as the sparse Pauli operator (SPO) and denote it informally as $\{(P_i, a_{P_i})\}$\fc{, where the index $i$ labels the Pauli strings retained in the stored sparse list rather than the full Pauli basis.}

The Pauli propagation simulation, or SPD, is a method that estimates the expectation value of observables $\langle 0 | U^\dagger O U |0\rangle$ under the unitary dynamics $U$ by time evolving the operator $O$ as a SPO.
The unitary $U$ has the form of a quantum circuit,
\begin{equation}
U=U_{\sigma_T}(\theta_T)\cdots U_{\sigma_1}(\theta_1), \qquad U_\sigma(\theta) = \exp(-i\theta \sigma / 2)
\end{equation}
where $\sigma$ is a Pauli operator defining the unitary rotation $U_\sigma(\theta)$ with angle $\theta$.
The SPD simulation consists of three components:
\begin{enumerate}
    \item Expressing the initial observable $O(0)$ at $t=0$ as a SPO
    $\{ (P^{[0]}_i,  a^{[0]}_{P_i} ) \}$
    \item The update rule to propagate the SPO $\{(P^{[t]}_i,  a^{[t]}_{P_i} ) \}$ through the $(t+1)$-th gate: 
    \begin{equation*}
        U^\dagger_{\sigma_{t+1}} O(t) U_{\sigma_{t+1}} = O(t+1)
    \end{equation*} 
    \begin{equation}
        \{(P^{[t]}_i,  a^{[t]}_{P_i} ) \} \longrightarrow 
        \{(P^{[t+1]}_i,  a^{[t+1]}_{P_i} ) \}
    \end{equation}
    \item Summation with the initial state, which typically is $| 0 \rangle \langle 0| = (({I+Z})/{2})^{\otimes N}$.
\end{enumerate}

The update rule is the essence of the simulation.
For any given Pauli rotation operator
\begin{equation}
U_\sigma(\theta) = \exp(-i\theta \sigma / 2) = \cos(\theta/2) -i\sin(\theta/2) \sigma ,
\end{equation}
we can separate the current list of Pauli strings $\{(P_i,  a_{P_i} ) \}$ into two sets $\mathcal{P}_\mathcal{C}$ and $\mathcal{P}_\mathcal{A}$.
The commuting set $\mathcal{P}_\mathcal{C}$ consists of Pauli strings $P_i$ that commute with $\sigma$, i.e., $[P_i, \sigma] = 0$.
The anti-commuting set $\mathcal{P}_\mathcal{A}$ consists of Pauli strings $P_i$ that anti-commute with $\sigma$, i.e., $\{P_i, \sigma\}=0$.

The central part of SPD is the equation
\begin{align} \label{eq: update_rule_1}
    &U^\dagger_\sigma(\theta) P U_\sigma(\theta) \nonumber \\
    &= (\cos(\theta/2) +i\sin(\theta/2) \sigma) P (\cos(\theta/2) -i\sin(\theta/2) \sigma) \nonumber \\
    &= \begin{cases}
        P, \qquad &\text{if}\  P \in \mathcal{P}_\mathcal{C} \\
        \cos(\theta) P + i \sin(\theta) \sigma\cdot P, \qquad &\text{if}\ P \in \mathcal{P}_\mathcal{A}  \\
    \end{cases}
\end{align}
which describes how one Pauli string evolves under a single gate $U_\sigma$ in the Heisenberg picture.
The update rule of the operator is then a combination of (i) branching: evolving all current Pauli strings with Eq.~\eqref{eq: update_rule_1} and (ii) merging: combining the coefficients of duplicated Pauli strings.
There are various ways to implement this update rule. The implementation usually has a strong dependency on the data structure that is used for storing the SPO.

We can rewrite Eq.~\eqref{eq: update_rule_1} to provide a different but equivalent formulation: %
With the given $U_\sigma(\theta), \mathcal{P}_\mathcal{C}, \mathcal{P}_\mathcal{A}$, we can further group Paulis inside the anti-commuting set $\mathcal{P}_\mathcal{A}$ in pairs $(P_i, Q_i)$, where the $Q_i$ is the conjugated Pauli string of $P_i$, defined by $\sigma \cdot P_i = i Q_i$.
If one member of the conjugated pair $P_i\ (Q_i)$ does not appear in the current set $\{(P_i,  a_{P_i} ), \  \forall P_i\in \mathcal{P}_\mathcal{A}\}$, we can always include it with a coefficient $a_{P_i}=0\ (a_{Q_i}=0)$.

With the above setup, we can now describe the update rules for $\mathcal{P}_\mathcal{C}$ and $\mathcal{P}_\mathcal{A}$. 
\begin{align}\label{eq: forward}
    \{(P^{[t]}_i,  a^{[t]}_{P_i} ) \} &\xrightarrow{U(\theta_{t+1})} 
    \{(P^{[t+1]}_i,  a^{[t+1]}_{P_i} ) \} \nonumber \\
    &\hspace{10mm}\mathclap{\rule{7cm}{0.4pt}}\nonumber \\[0mm]
     a^{[t]}_{P_i} &= a^{[t+1]}_{P_i}  , \qquad P_i \in \mathcal{P}_\mathcal{C} \nonumber \\[2mm]
     \mathcal{U}(\theta_{t+1})
     \begin{pmatrix}
      a^{[t]}_{P_i}\\[2mm]
      a^{[t]}_{Q_i}
     \end{pmatrix} &= 
     \begin{pmatrix}
     a^{[t+1]}_{P_i}\\[2mm]
     a^{[t+1]}_{Q_i}
     \end{pmatrix} ,
      \ (P_i, Q_i) \in\mathcal{P}_\mathcal{A}
\end{align}
where
\begin{equation}
    \mathcal{U}(\theta)= \begin{pmatrix}
     \cos(\theta) & \sin(\theta)\\[2mm]
     -\sin(\theta) & \cos(\theta)
     \end{pmatrix}
\end{equation} is a simple rotation over the $(P_i,Q_i)$ basis.
As a result, the action of a unitary gate and its conjugate on the observable is equivalent to applying a block diagonal rotation to the coefficient vector, where the block size is $2\times 2$.
This provides a physical picture of the update rule and explains why it is reversible and has a simple formula for the gradient.

At the end of the propagation, we can evaluate the physical observables by
\begin{align}
   \langle O \rangle &=  \text{Tr}[|0\rangle\langle 0| O(t)] = \text{Tr}\left[\left(\frac{I+Z}{2}\right)^{\otimes N} O(t) \right] \nonumber \\
   &= \sum_{P_i\in \mathcal{P}_{I,Z}} a_{P_i} ,
\end{align}
where $\mathcal{P}_{I, Z}$ denotes the set of Pauli strings composed only of $I$ and $Z$ operators.

\subsection{\label{subsec: bp} Backpropagation}
Next, we describe an efficient algorithm, both in runtime and memory, for obtaining the gradient to the same accuracy as the SPD simulation itself.
The main insight comes from utilizing the reversibility of the unitary circuit.
We can propagate the observables backward by propagating through the Hermitian conjugate circuit.
At the same time, this ability makes it possible to compute the approximate gradient in the sense of reverse-mode automatic differentiation.
Note that the difference from standard reverse-mode automatic differentiation~\cite{baydin2018automatic} is that no intermediate state is cached here.
This memory-efficient way of computing gradients is also the underlying idea utilized in libraries like Yao.jl~\cite{YaoFramework2019} and tensor network optimization~\cite{lin2021real,jobst2022finite,gibbs2025deep}.

\fc{
At a high level, the algorithm consists of one forward SPD sweep followed
by one backward reconstruction sweep, as illustrated in
Fig.~\ref{fig: illustration}.
The forward sweep propagates the initial SPO through the circuit and retains
only the terminal SPO.
At the terminal step, we initialize the coefficient adjoints
$\bar a_P\equiv\partial\mathcal{L}/\partial a_P$ from the cost function.
We then traverse the gates in reverse order, reconstructing the preceding operator coefficients with the inverse Pauli update, propagating the coefficient derivatives and accumulating the parameter gradients.
Consequently, only the current reconstructed SPO and its coefficient derivatives need to be stored.
The following subsections derive each of these steps.
}

\subsubsection{Evolving backward in time}

Consider a Pauli rotation operator
\begin{equation*}
U_\sigma(\theta) = \exp(-i\theta \sigma / 2) = \cos(\theta/2) -i\sin(\theta/2) \sigma\,.
\end{equation*}
The conjugated operator is
\begin{equation}
U^\dagger_\sigma(\theta) = \exp(+i\theta \sigma / 2) = \cos(\theta/2) +i\sin(\theta/2) \sigma\,.
\end{equation}
Now the update becomes
\begin{equation*}
    U_\sigma(\theta) P U^\dagger_\sigma(\theta)
    = \begin{cases}
        P, \qquad &\text{if}\ P \in \mathcal{P}_\mathcal{C} \\
        \cos(\theta) P - i \sin(\theta) \sigma\cdot P, \qquad &\text{if}\ P \in \mathcal{P}_\mathcal{A} \,.
    \end{cases}
\end{equation*}

In our notation, this naturally translates to the inverse of the original update rule.
Using the same notation as Eq.~\eqref{eq: forward}, we have
\begin{align}
\label{eq: backward}
    \{(P^{[t-1]}_i,  a^{[t-1]}_{P_i} ) \}& \xleftarrow{U^\dagger (\theta_{t})} 
    \{(P^{[t]}_i,  a^{[t]}_{P_i} ) \} \nonumber \\
    &\hspace{10mm}\mathclap{\rule{7cm}{0.4pt}}\nonumber \\[0mm]
    a^{[t-1]}_{P_i} &= a^{[t]}_{P_i}  , \qquad P_i \in \mathcal{P}_\mathcal{C} \nonumber \\
     \begin{pmatrix}
     a^{[t-1]}_{P_i}\\
     a^{[t-1]}_{Q_i}
     \end{pmatrix} &=
     \mathcal{U}(\theta_{t})^\dagger
     \begin{pmatrix}
      a^{[t]}_{P_i}\\
      a^{[t]}_{Q_i}
     \end{pmatrix},
      \ P_i\in\mathcal{P}_\mathcal{A}
\end{align}
and
\begin{equation}
\mathcal{U}^\dagger(\theta)=
    \begin{pmatrix}
     \cos(\theta) & -\sin(\theta)\\
     \sin(\theta) & \cos(\theta)
    \end{pmatrix}\,.
\end{equation}

In the following discussion, we use distinct notation for the coefficients $\tilde{a}_{\tilde{P}_i}$ and Paulis $\tilde{P}_i$ obtained in the backward path. In principle, the Paulis and coefficients will be slightly different between the forward and backward paths due to finite truncation error, except at the final step $t$, where $P^{[t]} = \tilde{P}^{[t]}$ and $\tilde{a}_{\tilde{P}_i} = {a}_{{P}_i}$ by definition. 
The forward and backward paths match exactly if the truncation threshold is set to zero. In this case, we obtain the numerically exact gradients.

\begin{figure}
    \centering
    \includegraphics[width=1.\linewidth]{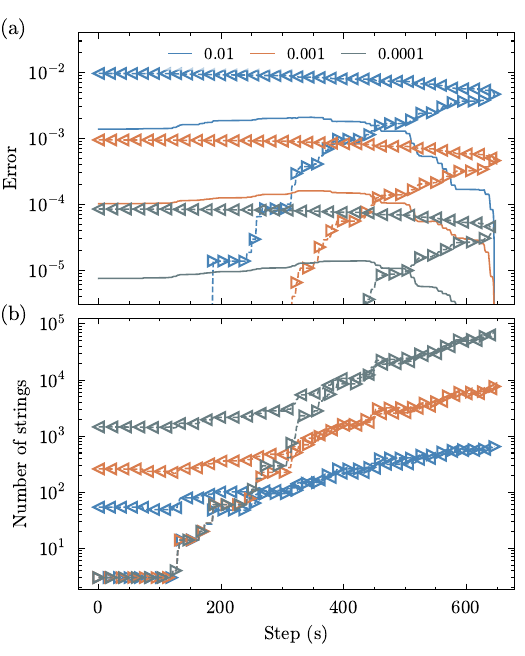}
    \caption{Error (a) and the number of strings (b) for forward $\triangleright$ and backward $\triangleleft$ propagation of the Hamiltonian operator with different truncation thresholds: $10^{-2}$ (blue), $10^{-3}$ (orange), and $10^{-4}$ (grey). 
    (a): The triangle markers track the truncation error in $l^2$ norm, i.e., $1 - \lVert O(s) \lVert / \lVert O(0) \rVert $ for both forward $\triangleright$ and backward $\triangleleft$ step $s$.
    The solid line indicates the error in overlap between operators from the forward and backward pass at the same step, i.e., $1 - |\langle O^\text{f}(s) | O^\text{b}(s)\rangle | / (\lVert O^\text{f}(s) \lVert  \lVert O^\text{b}(s) \lVert)$.
    (b): The number of Pauli strings at each step under forward and backward propagation. 
    }
 
    \label{fig: forward_backward_run}
\end{figure}

Below, we present a simple example to illustrate the effects of finite truncation error in the forward and backward propagation. 
We consider a depth-6 quantum circuit on an $8\times 8$ lattice with periodic boundary condition, which prepares an approximate ground state of a 2D transverse-field Ising model with $g=3.1$ near its critical point.  The detailed setup is not essential for the present discussion and is deferred to Sec.~\ref{sec: TFI_2D}. The key point is that the circuit under consideration is physically relevant and non-trivial.
The initial operator we choose is the combination of horizontal ZZ coupling, vertical ZZ coupling, and an onsite X term \fc{$O(0) = - Z_{ij}Z_{(i+1)j}  -Z_{ij}Z_{i(j+1)} - gX_i$}. The expectation value of this operator gives the energy density, so this is precisely the operator of interest.

We evolve the operator forward and backward with different truncation thresholds $\delta=10^{-2}$, $\delta=10^{-3}$, and $\delta=10^{-4}$.
The perfect evolution preserves the operator norm and yields exactly the same operator in the backward pass as in the forward pass at each step $s$.
Therefore, we monitor two types of errors: (i) the decay in the operator norm and (ii) the infidelity between the forward and backward operators.

Recall, with respect to the Pauli basis, the rescaled Hilbert-Schmidt inner product coincides with the standard $\ell^2$ inner product on the coefficients and the rescaled Hilbert–Schmidt (rHS) norm corresponds to the $\ell^2$ norm of the coefficients.
\begin{equation}
    \lVert O \rVert \coloneq \sqrt{\text{Tr}[O^\dagger O  / 2^n]} = \sqrt{ \sum_i |a_{P_i}|^2}\,.  
\end{equation}
The decay in the norm is given by
\begin{equation}
    1 - \frac{\lVert O(s) \lVert}{\lVert O(0) \lVert} = 1 - \frac{\sqrt{\langle \vec{a}(s)|\vec{a}(s)\rangle}}{\sqrt{\langle \vec{a}(0) |\vec{a}(0) \rangle }}\,,
\end{equation}
and the infidelity of operators between the forward and backward propagation is given by
\begin{equation}
    1 - \frac{|\langle O^\text{f}(s) | O^\text{b}(s)\rangle | }{ \lVert O^\text{f}(s) \lVert  \lVert O^\text{b}(s) \lVert} = 1 - \frac{\sqrt{\langle \vec{a}^f(s)| \vec{a}^b(s)\rangle \langle \vec{a}^b(s)| \vec{a}^f(s)\rangle } }{\sqrt{\langle \vec{a}^f(s)| \vec{a}^f(s)\rangle }\sqrt{\langle \vec{a}^b(s)| \vec{a}^b(s)\rangle }} .
\end{equation}
The superscripts ``f'' and ``b'' denote the forward and backward path respectively.

The result is shown in Fig.~\ref{fig: forward_backward_run}.
We see that the norm error in the forward propagation is of the same order of magnitude as that in the backward propagation. 
Despite the accumulation of the error as reflected by the loss of norm, the operator infidelity at each step $s$ remains roughly of the same order of magnitude throughout the full propagation. This indicates that the adjoint method can provide a memory-efficient evaluation of gradients without storing all the intermediate states.
We observe that the errors are roughly of the same order of magnitude of the truncation threshold $\delta$. However, this is not general and depends strongly on the underlying circuit. Since the error accumulates, a deeper circuit can easily have an order of magnitude larger error.
In general, the overall method is controlled, and we can reduce the error by adding cache points or using a smaller truncation threshold, at the cost of increased memory and runtime.

In Fig.~\ref{fig: forward_backward_run}(b), we plot the number of Pauli strings in the SPO. We observe that the number of strings peaks at the end of the forward propagation. The backward propagation follows the reverse of the forward propagation up to a point and then saturates at a finite value, which is a consequence of finite truncation leading to a slightly different path. This slight difference does not lead to an uncontrolled number of Pauli strings, and the backward SPO remains close to the forward one as monitored by the infidelity above.

\subsubsection{Gradient}

Having defined both forward and backward propagation, we are now in a position to show how to obtain the partial derivatives of the cost function with respect to the circuit parameters, i.e., the gradient.

Given an application and problem of interest, the cost function $\mathcal{L}$ quantifies the ``error'' of the solution associated with the parameters $\boldsymbol{\theta}$. The goal of the optimization is then to find the optimal solution  
\begin{equation}
    \text{argmin}_{\boldsymbol{\theta}}\  \mathcal{L}(\boldsymbol{\theta}) .
\end{equation}
In general, the cost function can be any differentiable function
\begin{equation}
    \mathcal{L}=\mathcal{L}(\{\rho_k\}, \{O_k\}, U(\boldsymbol{\theta}) )
\end{equation}
that depends on a set of initial operators $\{O_k\}$, initial states $\{\rho_k\}$, and the parameterized quantum operations $U(\boldsymbol{\theta})$, and can be efficiently evaluated with the sparse Pauli representation framework.
In this work, we focus on the parameterized quantum circuit with unitary gates, while it is possible to include more general operations, e.g., reversible quantum channels.

For simplicity, we further assume the cost function is a differentiable function of only the SPO at the final layer, $\mathcal{L}=\mathcal{L}(\{(a_{P^{[t]}_i}, P^{[t]}_i)\})$. This can be easily generalized to the dependency on multiple layers.
The backpropagation framework is agnostic to the cost function chosen.
The difference between cost functions only affects the initial values at the final layer, i.e., $\partial \mathcal{L} / \partial a^{[t]}_{P_i}$, for the backpropagation.

Starting from the initial vector ${\partial \mathcal{L}}/{\partial a^{[t]}_{P_i}}$, the chain rule gives the update rule for backpropagating the derivatives and computing the gradient with respect to the circuit parameter.
\begin{align}
    \frac{\partial \mathcal{L}}{\partial \tilde{a}^{[t-1]}_{P_j}} &= \sum_{i} \frac{\partial \mathcal{L}}{\partial \tilde{a}^{[t]}_{P_i}} \frac{\partial \tilde{a}^{[t]}_{P_i}}{\partial \tilde{a}^{[t-1]}_{P_j}} \label{eq: back-prop-1}\\
    \frac{\partial \mathcal{L}}{\partial \theta_{t}} &= \sum_i \frac{\partial \mathcal{L}}{\partial \tilde{a}^{[t]}_{P_i}} \frac{\partial \tilde{a}^{[t]}_{P_i}}{\partial \theta_{t}} \label{eq: back-prop-2}\,.
\end{align}
The first equation propagates the coefficient derivatives backward, while the second gives the parameter gradient with respect to $\theta_t$.

From the discussion in the previous section, at each step there are only very limited (one or two) contributions from $\tilde{a}^{[t-1]}_{P_j}$ to $\tilde{a}^{[t]}_{P_i}$.
More precisely, the Jacobian for the commuting set $\mathcal{P}_\mathcal{C}$ is the identity, ${\partial \tilde{a}^{[t]}_{P_i}}/{\partial \tilde{a}^{[t-1]}_{P_j}} = \delta_{ij}$, while the Jacobian for the anti-commuting set $\mathcal{P}_\mathcal{A}$ is block diagonal, ${\partial \tilde{a}^{[t]}_{P_{i,k}}}/{\partial \tilde{a}^{[t-1]}_{P_{j,l}}} = \delta_{ij}  \mathcal{U}_{k,l}$, where each block is given by the rotation $\mathcal{U}$. 
For notational simplicity, we suppress the notation $(P_i, Q_i) = (P_{i,1}, P_{i,2})$ here.

We can now complete the backward update rule:
\begin{align}
\label{eq: backward_gradient}
    \left \{(\tilde{P}^{[t-1]}_i,  \tilde{a}^{[t-1]}_{\tilde{P}_i}, \frac{\partial \mathcal{L}}{\partial \tilde{a}^{[t-1]}_{P_i}} ) \right \}&
    \xleftarrow{U^\dagger (\theta_t)}
    \left \{(\tilde{P}^{[t]}_i,  \tilde{a}^{[t]}_{\tilde{P}_i}, \frac{\partial \mathcal{L}}{\partial \tilde{a}^{[t]}_{P_i}} ) \right \}
    \nonumber \\
    &\hspace{5mm}\mathclap{\rule{7cm}{0.4pt}}\nonumber \\[0mm]
    \frac{\partial \mathcal{L}}{\partial \tilde{a}^{[t-1]}_{P_i}} &= \frac{\partial \mathcal{L}}{\partial \tilde{a}^{[t]}_{P_i}}  , \qquad \tilde{P}_i \in \mathcal{P}_\mathcal{C} \nonumber \\[2mm]
     \begin{pmatrix}
     \frac{\partial \mathcal{L}}{\partial \tilde{a}^{[t-1]}_{P_i}}\\[3mm]
     \frac{\partial \mathcal{L}}{\partial \tilde{a}^{[t-1]}_{Q_i}}
     \end{pmatrix} & = 
     \mathcal{U}^T (\theta_t)
     \begin{pmatrix}
     \frac{\partial \mathcal{L}}{\partial \tilde{a}^{[t]}_{P_i}}\\[3mm]
     \frac{\partial \mathcal{L}}{\partial \tilde{a}^{[t]}_{Q_i}}
     \end{pmatrix},
      \ \tilde{P}_i\in\mathcal{P}_\mathcal{A}
\end{align}
The transpose appears because the derivatives are propagated by the adjoint Jacobian.
We can easily extend the program shown in previous section to track the derivatives in addition to the coefficients in the backward pass.

Finally, to obtain the gradient with respect to the parameters $\theta$, we note that the contribution only comes from the anti-commuting set $\mathcal{P}_\mathcal{A}$, as the update for the commuting set $\mathcal{P}_\mathcal{C}$ does not depend on $\theta$.
From Eq.~\eqref{eq: forward}, we have for each pair $(\tilde{P}_i, \tilde{Q}_i)$
\begin{align*}
    \begin{pmatrix}
     \frac{\partial \tilde{a}^{[t]}_{\tilde{P}_i}}{\partial \theta_t}\\[3mm]
     \frac{\partial \tilde{a}^{[t]}_{\tilde{Q}_i}}{\partial \theta_t}
     \end{pmatrix}
     &= \frac{\partial \mathcal{U}(\theta_t)}{\partial \theta_t}
     \begin{pmatrix}
     \tilde{a}^{[t-1]}_{\tilde{P}_i}\\[2mm]
     \tilde{a}^{[t-1]}_{\tilde{Q}_i}
     \end{pmatrix} \\
     &= \frac{\partial \mathcal{U}(\theta_t)}{\partial \theta_t} \,\mathcal{U}(\theta_t)^\dagger
     \,\begin{pmatrix}
     \tilde{a}^{[t]}_{\tilde{P}_i}\\[2mm]
     \tilde{a}^{[t]}_{\tilde{Q}_i}
     \end{pmatrix} \\
     &=
     \begin{pmatrix}
     \tilde{a}^{[t]}_{\tilde{Q}_i}\\[2mm]
     - \tilde{a}^{[t]}_{\tilde{P}_i}
     \end{pmatrix}\,.
\end{align*}
Substituting this into Eq.~\eqref{eq: back-prop-2}, we obtain the following final expression
\begin{equation} \label{eq: gradient}
    \frac{\partial \mathcal{L}}{\partial \theta_{t}} = 
    \sum_{(\tilde{P}_i, \tilde{Q}_i) \in \mathcal{P}_\mathcal{A}}  \left( \frac{\partial \mathcal{L}}{\partial \tilde{a}^{[t]}_{\tilde{P}_i}}  \tilde{a}^{[t]}_{\tilde{Q}_i}
    -
    \frac{\partial \mathcal{L}}{\partial \tilde{a}^{[t]}_{\tilde{Q}_i}} \tilde{a}^{[t]}_{\tilde{P}_i}
    \right)\,.
\end{equation}
Note this expression only depends on the coefficient derivatives and coefficients at the current step $t$~\footnote{Here the sum is understood to include each conjugate Pauli pair once;
equivalently, an implementation may enumerate both orientations and divide
the result by two.}. Therefore, we only need to keep track of the derivatives and coefficients at a single time step, leading to a memory-efficient algorithm.

There are some subtleties whose discussion we deferred until now. This includes the truncation rule and non-sparse derivatives.
In this work, we store the derivative jointly with the SPO, treating the derivative as an extra value in addition to the coefficient.
We use the basic truncation rule based on the magnitude of the coefficients $|\tilde{a}_{P_i}|$. If the magnitude is smaller than the threshold $|\tilde{a}_{P_i}|<\delta$, we discard the tuple $(\tilde{P}_i,  \tilde{a}_{\tilde{P}_i}, {\partial \mathcal{L}}/{\partial \tilde{a}_{P_i}} ) $ including both the coefficient and the derivative.
From Eq.~\eqref{eq: gradient}, it is possible that the gradient contribution from $({\partial \mathcal{L}}{/\partial \tilde{a}_{\tilde{P}_i}} ) \tilde{a}_{\tilde{Q}_i}$ is not small even when
$|\tilde{a}_{P_i}|<\delta$.
However, we find the basic truncation rule works.
We envision that an improved truncation scheme targeting more accurate gradient is possible by taking a weighted sum based on $| \tilde{a}_{\tilde{P}_i}|$ and  $|{\partial \mathcal{L}}/{\partial \tilde{a}_{P_i}}|$.

The fact that the basic truncation rule works can be explained as follows. The derivative at the final layer $\partial \mathcal{L} / \partial a^{[t]}_{P_i}$ can often be non-sparse over the full Pauli basis.
Explicitly backpropagating the full derivative vector is therefore not feasible. Our basic truncation rule is to restrict the backward propagation only from the existing Pauli strings $|{a}_{P_i}|>\delta$.
We note that this approximation is distinct from truncating the coefficients of the derivative $\partial_{\theta_t}a_{P_i}=\pm a_{Q_i}$ that are smaller than $\delta$, because having a coefficient $a_{P_i}$ smaller than $\delta$ in the observable does not guarantee that $\partial_{\theta_t}a_{P_i}$ is also smaller than $\delta$. However, it is intuitively expected that a small $a_{P_i}$ correlates with a small $\partial_{\theta_t}a_{P_i}$, except for fine-tuned values of parameters $\theta$.
Indeed, having a large coefficient $a_{P_i}$ in the derivative with respect to $\theta$, but a coefficient $|a_{P_i}|<\delta$ in the observable, means that by modifying slightly $\theta$, one can make the coefficient $a_{P_i}$ in the observable larger than $\delta$. This implies that the parameters $\theta$ have to be fine-tuned for this situation to occur. As the truncation threshold is lowered, this situation becomes moreover less and less likely to occur. The bias introduced by the approximation can thus be systematically reduced by lowering the truncation threshold.

\subsubsection{Comparison with other differentiation schemes}
\label{subsec: diff_schemes}

The proposed backpropagation algorithm should be compared with several alternatives for evaluating gradients in an SPD simulation. These include the finite-difference methods and the auto-differentiation methods. 
For simplicity, in the following complexity comparison we assume that every parameterized gate carries an independent parameter, so that $\nparam$ is also the number of parameterized gates~\footnote{When parameters are shared, the gate-level derivatives are summed over all gates: the tangent dimension in forward mode is set by the number of independent parameters, whereas the storage required by a conventional cached reverse-mode implementation is set by the number of parameterized gates.}.

\begin{figure}[h]
    \centering
    \includegraphics[width=\linewidth]{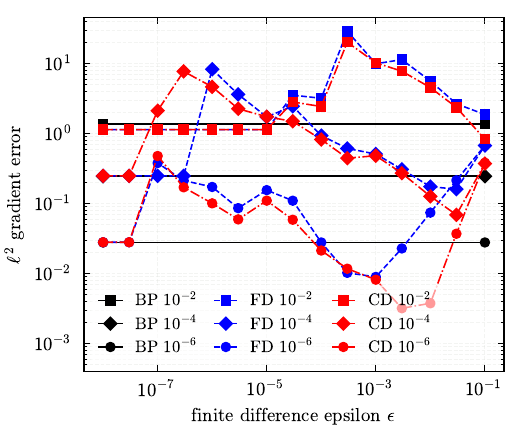}
    \caption{
    We compare the $\ell^2$ error of the gradient computed with the backpropagation (BP) method and two finite-difference methods using forward difference (FD) and central difference (CD), respectively.
    We consider the five-layer symmetry-breaking ansatz of
    Eq.~\eqref{eq: symm_breaking_ansatz} with 15 parameters for the 1D TFI model with $N=13$, $g=1.3$, and periodic boundary condition.
    The high-accuracy reference gradient is computed in double precision
    using BP with threshold $10^{-12}$.
    Markers denote truncation thresholds
    $\delta=10^{-2},10^{-4},10^{-6}$.
     }
    \label{fig: gradient_error}
\end{figure}

Finite-difference methods are memory efficient, since they only require repeated evaluations of the original observable.
However, they require $n_\text{param}$ additional function evaluations for forward differences and $2n_\text{param}$ for central differences, where $n_\text{param}$ is the number of circuit parameters.
Moreover, in the presence of finite truncation, the finite-difference step size $\epsilon$ introduces an additional numerical tradeoff: too large a value gives a biased derivative, an intermediate value can encounter the non-smoothness induced by finite truncation, and a small value risks the loss of precision.
Such sensitivity is illustrated in Fig.~\ref{fig: gradient_error}, where we compare the error in gradient estimates obtained from the proposed backpropagation method with forward- and central-difference methods.

Both forward-mode and reverse-mode automatic differentiation avoid the choice of a finite-difference step size and are efficient in runtime, as both require $\mathcal{O}(1)$ passes.
However, forward-mode automatic differentiation requires propagating the derivative of every retained coefficient with respect to every parameter.
This increases the memory footprint by a factor of $n_\text{param}$ compared with the original SPO representation; the corresponding forward-mode equations are given in Appendix~\ref{appendix: forward_mode}.
We note that the forward-mode also incurs an increase of $\mathcal O(\nparam)$ operations for coefficient update but this is typically the sub-leading term for the runtime cost.
Standard reverse-mode automatic differentiation implementations store the intermediate SPOs generated during the forward pass for the parameterized gates.
Its memory cost therefore scales again with the number of parameters $n_\text{param}$.

In contrast, the proposed backpropagation method avoids the need to tune $\epsilon$, computes all gradient components in one backward pass, and avoids the storage of intermediate SPOs by reconstructing the required intermediate operators during the backward sweep.
We summarize the memory and runtime comparison in Table~\ref{tab: comparison_diff}.
The runtime entries are stated in units of a single SPD expectation-value evaluation and ignore constant factors.
For the proposed method, the cost is a forward propagation followed by one backward propagation, so it remains $\mathcal{O}(1)$ with respect to $n_\text{param}$.

\begin{table}[htbp]
\centering
\begin{tabular*}{\columnwidth}{@{\extracolsep{\fill}} c c c c c }
 \hline
 Scheme & Forward-mode & Reverse-mode & finite diff. & This work\\ 
 \hline\hline
 Memory & $\mathcal{O}(n_\text{param} N_P)$ & $\mathcal{O}(n_\text{param} N_P)$ & $\mathcal{O}(N_P)$ & $\mathcal{O}(N_P)$\\
 Runtime & $\mathcal{O}(1)$ &  $\mathcal{O}(1)$ &  $\mathcal{O}(n_\text{param})$ &  $\mathcal{O}(1)$ \\
 \hline\hline
\end{tabular*}
\caption{ 
The summary table of the memory cost and the runtime of several methods to compute the gradient, including forward-mode automatic differentiation, standard reverse-mode automatic differentiation, finite-difference methods, and the method introduced in this work. We denote the number of parameters by $\nparam$ and the number of Pauli strings by $N_P$.
}
\label{tab: comparison_diff}
\end{table}

\section{\label{sec:results} Results}

In this section, we use the backpropagation algorithm to optimize quantum circuits for two applications: (i) low-energy state preparation and (ii) compression of time evolution circuits.
We consider two paradigmatic examples, the transverse field Ising (TFI) models 
\begin{equation}
    H_\text{TFI} = -\sum_{\langle i,j \rangle} Z_i Z_j - g\sum_i X_i
\end{equation}
and the anti-ferromagnetic Heisenberg (AFH) model,
\begin{equation}
    H_\text{AFH} = \frac{J}{4}\sum_{\langle i,j \rangle} \left( X_i X_j + Y_i Y_j + Z_i Z_j \right) .
\end{equation}
We study the TFI model on 1D, 2D, and 3D lattices, and use the AFH model as an additional 3D benchmark.
We set $J=1$ throughout.
All simulations use finite lattices with periodic boundary conditions (PBC); where a thermodynamic-limit value is reported, the lattice is chosen larger than the circuit light cone so that the local energy density is size independent.

\fc{
The optimization setup is the same throughout this section.
For a given circuit ansatz and cost function, we compute derivatives for all parameterized gates using the backpropagation method.
We further utilize the translational invariance over the PBC to reduce the computational complexity. 
We compute the cost and gradient using a few local terms.
Since the ansatz uses shared variational parameters, for example one angle for all single-site $X$ rotations in a layer, the gate-level derivatives are summed to obtain the gradient with respect to the shared parameter~\footnote{While we utilize the spatial average to reduce the computational complexity, the method applies equally well to non-translational invariant system.}.
Unless stated otherwise, the resulting cost and gradient are passed to L-BFGS-B with random or warm-started initial parameters.
The numerical accuracy is controlled by the truncation threshold, the maximum number of retained Pauli strings, and the floating-point precision.
Throughout the work, we estimate truncation-induced observable errors using the root-sum-square accumulation of the discarded $\ell^2$ weights discussed in Appendix~\ref{appendix: error}.
}

\subsection{Low-energy state preparation}
\label{subsec: result_state_prep}

The first application we consider is the preparation of low-energy quantum states.
It can be formulated as a minimization problem of the energy expectation value over the parameters in the given variational quantum circuit ansatz.
We consider a parameterized circuit $U(\boldsymbol{\theta})$ such that $|\psi(\boldsymbol{\theta})\rangle = U(\boldsymbol{\theta})|0\rangle$.
The energy expectation value is the cost function
$E = \langle \psi| \hat{H} | \psi \rangle = \langle 0| U(\boldsymbol{\theta})^\dagger \hat{H} U(\boldsymbol{\theta}) | 0 \rangle = \langle 0| \hat{H}(t; \boldsymbol{\theta}) | 0 \rangle $.
As discussed in Sec.~\ref{subsec: spd}, evaluating the expectation value in the $|0\rangle$ state amounts to taking the operator at the final step and summing over the coefficients in the $I,Z$ basis,
$\sum_{P_i\in \mathcal{P}_{I,Z}} a^{[t]}_{P_i}$.
The initial derivative is a Kronecker-delta vector over the $I,Z$ basis.
\begin{equation}
\frac{\partial\mathcal{L}}{\partial a^{[t]}_{P_i}} = \begin{cases}
    1  \qquad \textrm{if}\  P_i\in \mathcal{P}_{I,Z} \\
    0  \qquad \textrm{otherwise}
\end{cases} .
\end{equation} 
In the following, we perform optimization by propagating only the derivative with non-zero coefficient support.

\subsubsection{Transverse-field Ising model in 1D}

In 1D, the TFI model can be mapped to a free fermion model through Jordan-Wigner transformation and is exactly solvable.
The model has a quantum critical point at $g_c=1$. The ground states belong to a $Z_2$ symmetry broken ferromagnetic phase for $g<1$ and belong to a paramagnetic phase for $g>1$.

\begin{figure}[th]
\includegraphics[width=0.5\textwidth]{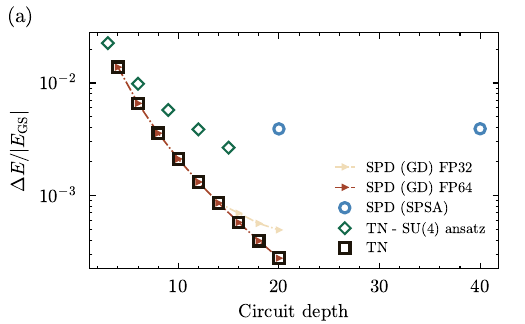}
\includegraphics[width=0.5\textwidth]{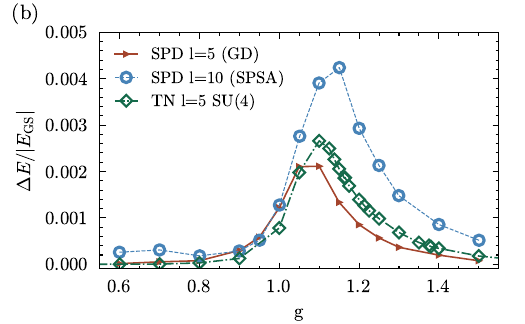}
\caption{
(a) Relative energy error of the 1D TFI $g=1.1$ against circuit depth in terms of entangling two-qubit gates for several optimization methods and ansatz choices. The first set of data points (green diamond) is a generic SU(4) brickwork ansatz optimized with tensor network methods~\cite{jobst2022finite}. The second (black square), third, and fourth set of data points use the HVA ansatz optimized by gradient descent, where the gradient is obtained through tensor network methods~\cite{sokolov2025bang} or through SPD backpropagation at two numerical precisions.
The last set of data points comes from the SPSA optimization of the generic ansatz including the symmetry-breaking term taken from~\cite{lin2026utility}.
(b) Relative energy error against different $g$. We show our gradient descent optimization result with the variational ansatz with Eq.~\eqref{eq: generic_ansatz} of 5 layers (depth-10).
We compare it with data from~\cite{lin2026utility} using the same ansatz optimized with SPSA and SPD of 10 layers (depth-20). We also include the data from~\cite{jobst2022finite} using generic SU(4) brickwork ansatz of 5 layers (depth-15) and optimized with tensor network methods.
}
\label{fig: 1d convergence}
\end{figure}

As an exactly solvable model, the 1D TFI model has been extensively studied in the literature for finite-depth scaling of the critical exponent~\cite{dreyer2021quantum,jobst2022finite} and quantum state preparation~\cite{sokolov2025bang,lin2026utility}.
It therefore provides a useful benchmark for the following two questions.
First, in a setting where SPD can be made exact, does the proposed backpropagation algorithm reproduce established gradient-based optimization results?
Second, once gradients are available, how much does gradient-based optimization improve over the stochastic approximated gradient approach previously used with SPD?

We answer these questions using two closely related variational ans\"atze. The more general ansatz includes a symmetry-breaking term~\cite{park2024efficient},
\begin{equation}\label{eq: generic_ansatz}
    U(\alpha,\beta,\gamma) = u_Z(\alpha)u_X(\beta)u_{ZZ}(\gamma)\,,
\end{equation}
where
\begin{align}
    u_{Z}(\alpha) &= \prod_{i} \exp(-i\frac{\alpha}{2}Z_i) \nonumber \\
    u_{X}(\beta) &= \prod_{i} \exp(-i\frac{\beta}{2}X_i) \nonumber \\
    u_{ZZ}(\gamma) &= \prod_{\langle i j \rangle} \exp(-i\frac{\gamma}{2}Z_i Z_j). 
\end{align}
Starting from the paramagnetic fixed point, the variational wavefunction of the $l$-layer is then given as
\begin{equation} \label{eq: symm_breaking_ansatz}
  |\psi (\boldsymbol{\alpha}, \boldsymbol{\beta}, \boldsymbol{\gamma})  \rangle  = U(\alpha_l,\beta_l,\gamma_l)\cdots U(\alpha_1,\beta_1,\gamma_1) | + \rangle^{\otimes N}\,.
\end{equation}
This ansatz was considered in Ref.~\cite{park2024efficient,lin2026utility} as a generalization of the standard Hamiltonian variational ansatz (HVA)~\cite{dreyer2021quantum,sokolov2025bang}.
If we restrict all $\alpha_l=0$ in Eq.~\eqref{eq: symm_breaking_ansatz}, we recover the standard HVA 
\begin{align}
\label{eq: HVA}
U(\beta_l,\gamma_l) &= u_X(\beta_l)u_{ZZ}(\gamma_l) \nonumber \\
|\psi_\text{HVA} (\boldsymbol{\beta}, \boldsymbol{\gamma})  \rangle  &= U(\beta_l,\gamma_l)\cdots U(\beta_1,\gamma_1) | + \rangle^{\otimes N} .
\end{align}
After proper optimization, the standard HVA starting from the symmetric product state $|+\rangle^{\otimes N}$ converges exponentially with circuit depth in the paramagnetic (symmetric) phase and slows down to algebraic convergence when crossing the critical point~\cite{dreyer2021quantum}. 
The generic symmetry-breaking ansatz is designed to restore exponential convergence throughout the phase diagram, except at the critical point~\cite{park2024efficient}.

We first restrict to the HVA case without the additional $u_Z$ layer.
This case is especially useful as a validation test because circuits generated solely by $u_X$ and $u_{ZZ}$ are fermionic Gaussian (free-fermion) circuits.  
Both $X_i$ and nearest-neighbor $Z_iZ_{i+1}$ are Majorana bilinears. Fermionic Gaussian evolution preserves the space of quadratic Majorana operators, whose dimension is \(\binom{2N}{2}=\mathcal O(N^2)\). Since every Majorana bilinear maps to a Pauli string, the propagated local Hamiltonian has at most \(\mathcal O(N^2)\) Pauli support, so the SPD calculation remains exact with polynomial resources.
The same statement applies to the backward propagation.
As a result, the SPD computation of both expectation values and gradients is exact up to floating-point error.
Thus, this benchmark tests the gradient algorithm itself without the additional complication of truncation error.

It also allows a direct comparison with the tensor network optimization results from Ref.~\cite{sokolov2025bang}, which are also numerically exact~\footnote{There are several other possible benchmarks. For example, closed-form expressions for the energy expectation value are available~\cite{dreyer2021quantum}, and for the depths considered here the light cone is small enough for exact statevector simulation. For simplicity, we compare directly with existing data from the literature.}.
We optimize the HVA with increasing number of layers $l$ at $g=1.1$.
We consider PBC systems with system size $N > 2(l+1)$, which is larger than the light cone of the evolved operator. As a result, the optimized energy density can be compared directly with the ground state energy density in the thermodynamic limit $E_\text{GS}$.

In Fig.~\ref{fig: 1d convergence}(a), we plot the relative error in the energy density $(E_l - E_\text{GS})/|E_\text{GS}|$ against the number of layers $l$.
The raw energy densities are listed in Appendix~\ref{appendix: energy}.

We observe agreement with the tensor network data to all reported digits when double precision is used.
In single precision, the optimization deviates from the best result at larger depth because of precision loss in the gradient evaluation. We provide discussion and testing on this effect in Appendix~\ref{appendix: numerical_precision}.
We also show the previous SPD result of Ref.~\cite{lin2026utility}, obtained with SPSA and the symmetry-breaking ansatz of Eq.~\eqref{eq: symm_breaking_ansatz}.
Since that curve uses a different ansatz from the HVA data, we regard it as a reference for prior SPD optimization rather than as a direct HVA benchmark~\footnote{Ref.~\cite{lin2026utility} considers a finite PBC system of size $N=100$. The comparison is valid for two reasons. Firstly, the finite ground-state correlation length, $\xi\sim 10.4$~\cite{sokolov2025bang}, renders the energy density of a finite size system indistinguishable from the thermodynamic limit. Secondly, the circuit light cone remains smaller than the system size, ensuring that the energy density measured remains invariant with increasing system size.}.
We also include data from the study of infinite brickwork circuit~\cite{jobst2022finite}.
The resulting errors are comparable despite the fact that the brickwork ansatz consists of generic SU(4) two-qubit gates.
To make a fairer comparison, we compare the accuracy in terms of the depth of elementary two-qubit entangling gates, as each generic SU(4) gate requires $3$ two-qubit entangling gates.
Overall, the HVA benchmark validates the backpropagation algorithm in a setting where the only numerical limitation is floating-point precision.

We next turn to the optimization of generic symmetry-breaking ansatz and vary the transverse field $g$ across both phases.
The result is shown in Fig.~\ref{fig: 1d convergence}(b).
Our final energy accuracy with the 5-layer ansatz reaches a similar, if not better, accuracy compared to the optimization result with the 10-layer ansatz using the SPSA method~\cite{lin2026utility} over the full phase diagram. 
This is the direct comparison with previous SPSA-based SPD optimization, since both calculations use the same symmetry-breaking ansatz.
This comparison provides evidence that having access to the gradient information enables more efficient and accurate optimization.
In addition, we compare with the generic 5-layer SU(4) ansatz optimized by tensor network methods and gradient descent algorithms. Despite the differences in ansatz structure and depth, namely depth 15 for the SU(4) brickwall ansatz and depth 10 for the 5-layer symmetry-breaking ansatz, the optimized energies have similar accuracy.

\subsubsection{Transverse-field Ising model in 2D
\label{sec: TFI_2D}
}

\begin{figure}[ht]
\includegraphics[width=0.5\textwidth]{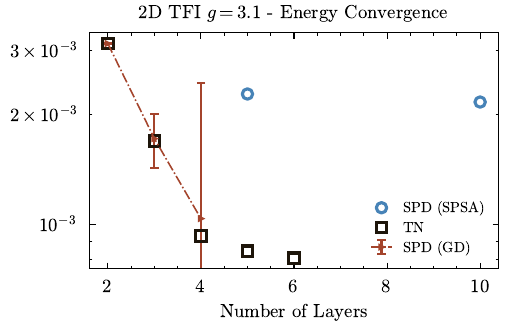}
\caption{
Relative energy error of the 2D TFI $g=3.1$ against the number of repeated layers in the variational ansatz.
We compare the optimization of the HVA circuit using SPD backpropagation based gradient descent (red) and the iPEPS-based gradient descent method~\cite{zhang2024bang} (black).
We also include data taken from~\cite{lin2026utility} with the generic ansatz in Eq.~\eqref{eq: generic_ansatz} optimized by SPSA and SPD on a $(10, 10)$ PBC lattice.
}
\label{fig: 2d convergence}
\end{figure}

After benchmarking with exact simulation in 1D, we turn to the 2D simulation where the SPD includes finite truncation.
We find the low-energy state preparation circuit for 2D TFI with $g=3.1$ near the critical point by minimizing the energy expectation value.
Here we consider the Hamiltonian variational ansatz as in Eq.~\eqref{eq: HVA} without the symmetry breaking terms.
We compare the result with infinite projected entangled-pair state (iPEPS)-based gradient descent optimization~\cite{zhang2024bang} and the previously reported SPD result using SPSA method~\footnote{The SPD--SPSA data of Ref.~\cite{lin2026utility} were obtained and evaluated on a finite $10\times10$ PBC lattice, and their relative errors use the corresponding finite-size reference energy. The SPD-backpropagation and tensor-network data instead use the iPEPS thermodynamic-limit reference energy. %
}.
We show the result in Fig.~\ref{fig: 2d convergence}.

\fc{For the first few layers, the SPD-optimized HVA reaches an energy density close to the iPEPS-optimized result, while using the same physically motivated ansatz.
At larger depth, the finite number of Pauli strings and the associated truncation error become effective. Currently, we cannot obtain a meaningful bound on the energy error for circuits with more than 5 layers.
This behavior highlights the main practical difference between the 1D and 2D benchmarks: in 1D the benchmark isolates the gradient algorithm, whereas in 2D the optimization also tests whether the backpropagation with finite truncation approximation remains sufficiently accurate for optimization.
The raw energy densities and the tracked SPD error estimates are reported in Appendix~\ref{appendix: energy}.}

\subsubsection{Three-dimensional models}

\fc{
Finally, we apply the same SPD backpropagation workflow to 3D lattices.
The purpose of these calculations is not to establish state-of-the-art variational energies, but to show that the method can easily be generalized to different geometries.
For the 3D TFI model at $g=5.2$, close to the critical point, a two-layer HVA circuit on an $8\times8\times8$ lattice with PBC reaches energy density $E=-5.3531$. The empirical SPD truncation-error estimate is $2.0\times10^{-4}$.
This is close to the reference quantum Monte Carlo (QMC) value $E=-5.35872(8)$ for the same finite lattice.
For the 3D AFH model, optimization on a $4\times4\times4$ lattice with PBC gives energy density $E=-0.889315$.
Evaluating the same circuit parameters on systems larger than the circuit light cone, $L\geq 8$, gives the effectively thermodynamic-limit energy density $E=-0.8891$. We can compare it with the QMC extrapolated estimate $E=-0.902325(11)$~\cite{vlaar2021simulation}.
All the corresponding numerical data are collected in Appendix~\ref{appendix: energy}.
The finite-size QMC values reported here were obtained using
ALPS~\cite{bauer2011alps}; the thermodynamic-limit AFH value is taken
from Ref.~\cite{vlaar2021simulation}.
}

\subsection{Time evolution compression}

The second application we consider is the compression of time evolution circuits.
The goal is to find a shallower quantum circuit $U(\boldsymbol{\theta})$ to approximate a deeper target quantum circuit $V$.
A typical cost function for this task is the Hilbert-Schmidt test cost $\mathcal{L}_\text{HST}$
\begin{equation}
    \mathcal{L}_\text{HST} = 1 - \frac{1}{4^n} \left \lvert \text{Tr}(U^\dagger V)  \right \rvert^2 .
\end{equation}
Here, instead, we use a more SPD-friendly cost function defined as the sum over the local X errors and Z errors
\begin{equation}
    \mathcal{L}_{X,Z} = \sum_i \left \lVert  X_i(t)  - \tilde{X}_i(t) \right\rVert^2_\text{rHS} + \sum_i \left \lVert  Z_i(t)  - \tilde{Z}_i(t) \right\rVert_\text{rHS}^2  
\end{equation}
where $\tilde{X}_i(t)$ and $\tilde{Z}_i(t)$ are the time-evolved $X_i$ and $Z_i$ generators with the target unitary $V$.
Recall, the rescaled Hilbert–Schmidt (rHS) norm corresponds to the $\ell^2$ norm of the coefficients.
If $\tilde{X}_i(t)$ and $\tilde{Z}_i(t)$ permit efficient sparse Pauli representation, then the cost function and the corresponding gradient can be easily evaluated from the difference between the sparse operators.
We note that this is a simplified construction of the previous work~\cite{d2025circuit}, where the slightly different local error cost function has been shown to both lower and upper bound the $\mathcal{L}_\text{HST}$.
We present a simplified derivation of both the lower and upper bound in the Appendix~\ref{sec: local cost}.

\subsubsection{Transverse-field Ising model in 2D}

Below, we show a prototype test on the variational time evolution compression task.
The setup is as follows.
We consider a square lattice of size $(10, 10)$ with PBC and the target continuous time evolution unitary $V=e^{-i\hat{H}T}$, where $\hat{H}$ is the 2D TFI with $g=3.1$ and total time $T=0.3$.
We can estimate the light cone size by checking the size of the Pauli strings. The system size is chosen to be larger than the light cone.

We consider the variational ansatz circuit $U(\boldsymbol{\theta})$ consisting of alternating layers of $U_X$ and $U_{ZZ}$. A circuit of $l$-layer $U_X(\theta_1) U_{ZZ}(\theta_2) \ldots U_X(\theta_{2l+1})$ has in total $2l+1$ independent parameters and a total depth $4l$ in terms of elementary entangling gates. 
The parameters are initialized with random numbers drawn from the uniform distribution $[-0.05, 0.05]$.
As both the ansatz circuit and the target unitary are single-site translationally invariant, the cost function is further simplified to the combination of one $X$ error and one $Z$ error.

We approximate the target unitary $V$ by a 4th-order Trotterization with $dt=0.001$ and truncation threshold $10^{-8}$ in the SPD simulation to obtain the target $X^\text{target} = V^\dagger XV$ and $Z^\text{target}= V^\dagger ZV$. We observe a loss in the norm of the two operators $\sim 2.26\times10^{-5}$ and $\sim 4.79\times 10^{-5}$, respectively.
The evaluations below with SPD are based on the same truncation threshold and reach similar accuracy.

\begin{figure}[th!]
    \centering
    \includegraphics[width=\linewidth]{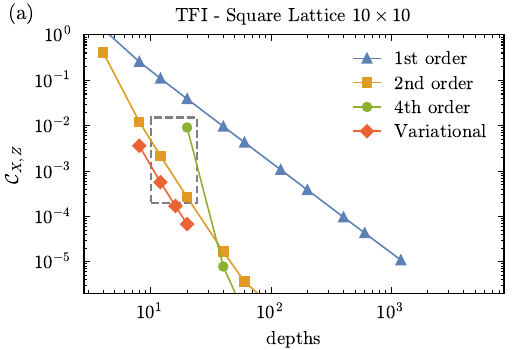}    \includegraphics[width=\linewidth]{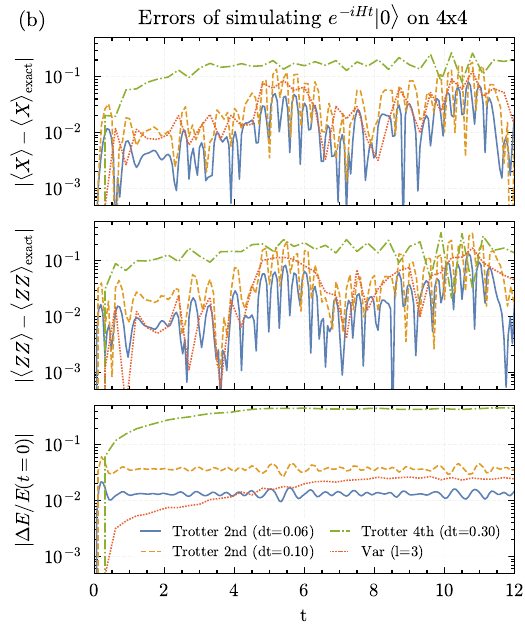}
    \caption{Time evolution circuit compression with 2D TFI model.
    (a) The $\mathcal{L}_{X,Z}$ cost of the variational compressed circuits with layers $l=2,\ldots,5$ compared to 1st-, 2nd-, and 4th-order Trotterized circuits of various $dt$. To have a fair comparison, we plot the cost against the depths of elementary 2-qubit entangling gates.
    (b) Errors of simulating the quantum quench  $|\psi(t)_\text{exact}\rangle = e^{-iHt}|0\rangle$ with the variationally compressed circuit compared to Trotterized circuits.
    We consider four different circuit constructions, including (i) 2nd-order Trotter with $dt=0.06$, (ii) 2nd-order Trotter with  $dt=0.1$, (iii) 4th-order Trotter with $dt=0.3$, and (iv) variational circuit with 3 layers.
    These four options correspond to the four data points inside the grey dashed box in (a).
    In terms of representing a time evolution circuit with $T=0.3$, (i) and (iii) require 5 layers of $U_{ZZ}$ while (ii) and (iv) require 3 layers.
    We compute errors of local expectation value $\langle X\rangle$, nearest neighbor correlation $\langle ZZ\rangle$, and the energy $E$ with respect to the exact continuous time evolution.
    }
    \label{fig: time_evolution_compression_tfi_2d}
\end{figure}

The result is shown in Fig.~\ref{fig: time_evolution_compression_tfi_2d}(a), where we compare the resulting cost functions of the variational circuit ansatz to standard Trotterized time-evolution circuits of order $1$, $2$, and $4$.
We observe that we can obtain the expected scaling relationship of different orders of Trotterization when the values of the cost functions are above the simulation accuracy $\sim 10^{-5}$.
The variational circuit ans\"atze with layers from $2$ to $5$ all reach lower cost function values compared to the Trotterized circuits.
We find that random initialization only works for layer-$2$ and layer-$3$. The optimization of deeper circuits requires initialization with values from Trotterized circuits.
Overall, we find that the reduction in the cost function is not significant. There are possibly various reasons. It could be due to the simplicity of the model such that the Trotter formula already gives a very compact circuit. Other possible reasons include the small total time $T$ and the restricted variational ansatz, which has strictly the same form as the Trotterized circuits.

We consider a concrete example of simulating quantum quench dynamics with circuits where the exact continuous time evolution is described by $|\psi(t)_\text{exact}\rangle = e^{-iHt}|0\rangle$.
We perform the state vector simulation on a $4 \times 4$ square lattice with PBC and compare the variational compressed circuit with 3 layers with 2nd-order Trotterized circuits with $dt=0.06,\ dt=0.1$ and a 4th-order Trotterized circuit with $dt=0.3$.
We compute errors of local expectation value $\langle X\rangle$, nearest neighbor correlation $\langle ZZ\rangle$, and the energy $E$ with respect to the exact continuous time evolution.
The result is shown in Fig.~\ref{fig: time_evolution_compression_tfi_2d}(b).

Among all circuits, the 2nd-order Trotterized circuit with $dt=0.06$ has the lowest $\mathcal{L}_{X,Z}$ error and gives overall consistently lower error throughout the time evolution.
It requires 5 layers of $U_{ZZ}$. In contrast, we see the 3-layer variational compressed circuit can obtain a comparable accuracy in local observables while having a better conservation in energy before $t=4$.
Since each layer of $U_{ZZ}$ corresponds to a depth-4 circuit in terms of elementary 2-qubit entangling gates, $t=4$ corresponds to $\sim 150$ depth using layer-3 circuits and to $\sim 250$ depth using layer-5 circuits.
These depths sit at the boundary of the limit of the current quantum device.
As a result, the ability of $40\%$ reduction in circuit depths is of much relevance to the current circuit design.

While the results in Fig.~\ref{fig: time_evolution_compression_tfi_2d}(b) are instructive, we emphasize that the quality of the results strongly depends on the chosen initial state.
Indeed, the error measure performs an average over the full Hilbert space in the sense that $\mathcal{L}_{X, Z}$ can be seen as the sum of the thermal expectation values of the error on the local $X$ and $Z$ at infinite temperature.
Therefore, while the optimized unitary should be closer to the continuous unitary in terms of the Frobenius norm, there is no guarantee that it will perform better than the Trotter unitary for all choices of initial states, as already noted in Ref.~\cite{Mansuroglu_2023}.
In our simulations, we have noticed that the performance of the compressed circuit was better when the initial state was closer to the center of the spectrum rather than at low energy.
This is consistent with the observation that the non-optimized Trotter circuit performs particularly well at low energies~\cite{q87n-5xhz}.

\section{Discussion}
\label{sec: discussion}

\subsection{Complexity and regularization}
\label{subsec: complexity-and-regularization}

The computational complexity of SPD simulation depends on both the data structure used to represent the SPO and its rank, i.e., number of Pauli strings $N_P$ in the decomposition.
In our implementation, the SPO is stored as a sorted array.
Each gate application requires searching for existing strings, updating coefficients, merging duplicate terms, and re-sorting the array, leading to a cost of $\mathcal{O}(N_P\log N_P)$ operations.

While the SPO rank provides a direct measure of simulation cost, it does not account for the distribution of coefficients. A more refined family of complexity measures for SPOs is the operator stabilizer R\'enyi entropy (OSE)~\cite{dowling2025magic}.
For a normalized operator $\hat{O}=\sum_P a_P P$ with $\sum_P |a_P|^2 = 1$, the OSE is defined as the participation R\'enyi-$\alpha$ entropy
\begin{equation}
    \mathcal{M}^{[\alpha]}(\hat{O}) = \frac{1}{1- \alpha} \log (\sum_{P\in\mathcal{P}} |a_{P}|^{2\alpha})\,,
\end{equation}
over the probability distribution $|a_P|^2$ on Pauli strings.
For $\alpha=1$, the corresponding expression is obtained by taking the Shannon limit,
$\mathcal{M}^{[1]}(\hat{O})=-\sum_P |a_P|^2\log |a_P|^2$.
The SPO rank is included in this family of measures in the limit $\mathcal{M}^{[\alpha\rightarrow 0]}=\log(\mathrm{rank})$.
In general, the OSE measures how spread out the distribution is and controls the truncation error in analogy with the role of entanglement entropy in matrix-product states methods.
Specifically, logarithmic scaling of R\'enyi entropies with $\alpha<1$ implies  efficient approximability by keeping only a polynomial number of coefficients~\cite{verstraete2006matrix,schuch2008entropy}.
Beyond these numerical implications, the OSE is a magic monotone and satisfies several desirable properties from the perspective of the magic resource theory, including faithfulness, stability, additivity, and boundedness~\cite{dowling2025magic}.

Regularization is a commonly used technique in optimization, where the cost function is augmented by a penalty term to control the complexity of the solution and hence, in various contexts, to prevent overfitting.
In the present setting, since the OSE provides an operationally meaningful complexity measure, a natural choice is to use it as the regularizer~\footnote{Other possible choices include $|\boldsymbol{\theta}|^2$ or the distance to a Clifford point.}
\begin{equation}
    \mathcal{L}_\text{total} = \mathcal{L} + \lambda_\text{OSE}\ \mathcal{M}^{[\alpha]} ,
\end{equation}
where $\lambda_\text{OSE}$ is the hyperparameter controlling the strength of the regularization.
The effect of the regularizer depends on the R\'enyi index $\alpha$.
Roughly speaking, R\'enyi entropies with $\alpha<1$ are more sensitive to the small-coefficient tail, whereas those with $\alpha>1$ are dominated by the largest coefficients.

The OSE regularization can be useful for SPD optimization in two complementary settings.
First, it provides a practical tool to reduce uncertainty when the simulation is restricted to a finite maximum number of Pauli strings $N_P$.
With finite truncation, direct minimization of $\mathcal{L}$ often produces circuits with apparently lower cost but larger truncation uncertainty.
These lower values of $\mathcal{L}$ can be artifacts of the approximation and may disappear when the same circuit is evaluated at higher accuracy.
The relevant figure of merit is therefore not only the estimated value of $\mathcal{L}$, but also its uncertainty, or more conservatively an upper confidence bound on $\mathcal{L}$.
Although the heuristic error estimate discussed in Appendix~\ref{appendix: error} is useful in practice, rigorous error bounds are often too loose to guide the optimization directly.
The optimization with regularization provides a practical way to address this issue. By tuning the regularization strength, one can favor less complex, more compressed operators, thereby reducing the uncertainty coming from the truncation at the cost of some variational performance.

Second, OSE regularization provides a way to probe the tradeoff between energy and non-stabilizerness.
In fault-tolerant quantum computation, non-Clifford gates, and in particular T gates, are costly resources.
Although the OSE gives only a lower bound on the required T-gate count, it quantifies the non-stabilizerness generated in the Heisenberg-evolved operators and therefore serves as a useful proxy for circuit magic.
As a result, SPD with OSE regularization thus offers a natural framework for studying how much non-stabilizerness is required to achieve a given energy accuracy.

\subsubsection{Regularization on 1D TFI circuits}

\begin{figure}[th]
\includegraphics[width=0.5\textwidth]{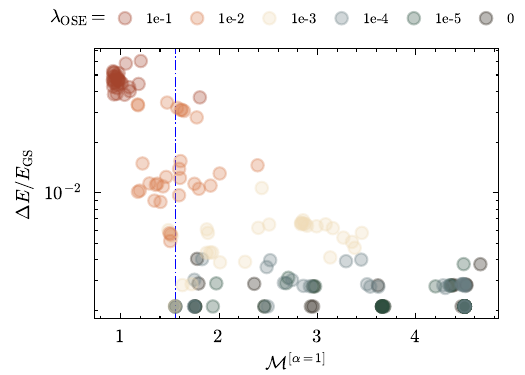}
\caption{
Relative energy error and operator stabilizer entropy $\mathcal{M}^{[\alpha=1]}$ of the circuits optimized with cost function $\mathcal{L}_\text{total} = E + \lambda_\text{OSE}\,\mathcal{M}^{[\alpha=1]}$. Different colors correspond to different regularization strength $\lambda_\text{OSE}$.
The dashed blue line indicates the smallest value of $\mathcal{M}^{[\alpha=1]}$ among circuits that reach the lowest observed energy.
}
\label{fig: regularization}
\end{figure}

We demonstrate the effect of OSE regularization by revisiting the task of finding low-energy state preparation circuit for the 1D TFI model for $g=1.1$.
The setup is as follows. We optimize the generic symmetry-breaking ansatz of Eq.~\ref{eq: symm_breaking_ansatz} with the cost function $\mathcal{L}_\text{total} = E + \lambda_\text{OSE}\,\mathcal{M}^{[\alpha=1]}$. We consider five-layer circuits and optimize over a system size $L=12$ larger than the light cone.
For each value of the regularization strength $\lambda_\text{OSE}$, we optimize over 30 random circuit realizations with parameters initialized uniformly $[-0.1,0.1]$.

The results are shown in Fig.~\ref{fig: regularization}, where we plot the relative energy error against the OSE of the propagated local Hamiltonian
$h_i = -Z_i Z_{i+1} - g X_i$.
We observe two qualitatively distinct regimes.
For stronger regularization, $\lambda_\text{OSE}\geq 10^{-3}$, the optimized circuits exhibit a clear tradeoff between energy accuracy and OSE: increasing $\lambda_\text{OSE}$ lowers the OSE, but at the cost of a larger energy error.
This behavior confirms that the regularizer can be used to bias the optimization toward lower-complexity, more compressible operators.
From the perspective of finite-budget SPD simulations, such lower-OSE solutions are expected to have less weight in the truncated tail at fixed $N_P$, and hence smaller truncation uncertainty.
Thus, the numerical results illustrate the first use case described above: OSE regularization can trade a controlled loss in variational accuracy for improved robustness under truncation.
At the same time, the same tradeoff can be interpreted as an energy--non-stabilizerness tradeoff, since the OSE quantifies the magic of the propagated operator.
This illustrates the second use case: varying $\lambda_\text{OSE}$ provides a systematic way to explore how much non-stabilizerness is required to reach a given energy accuracy.

For weaker regularization, $\lambda_\text{OSE}\leq 10^{-4}$, the optimized circuits reach similar energy errors but exhibit a broad spread in OSE values.
We interpret this behavior as a consequence of the large freedom in the state-preparation problem, although a complete explanation remains open.
One source of this freedom is that the variational objective only constrains the action of the circuit on the reference state,
$ U|0\rangle \approx |\psi_\text{low-energy}\rangle $
and therefore fixes only one column of the full unitary $U$.
Different circuits can therefore prepare nearly the same low-energy state while acting very differently on the rest of Hilbert space, resulting in large differences in the OSE of the Heisenberg-evolved operators.

A second, more subtle possibility is that the circuits prepare different states that have nearly identical energy expectation values but differ by relative phases in the energy eigenbasis.
For example, states of the form
$|\psi\rangle = c_0|\psi_\text{GS}\rangle + c_1 e^{i\theta}|\psi_\text{excited}\rangle$
have the same energy expectation value for different values of $\theta$, but the corresponding preparation circuits may induce substantially different Heisenberg evolutions and hence different OSE values.
These two mechanisms are not specific to the generic symmetry-breaking ansatz.
Indeed, we observe a similar broad spread of OSE values in the free-fermion HVA, suggesting that the effect is more closely tied to the freedom of state preparation than to the choice of ansatz.
Clarifying the relative importance of these mechanisms, and their dependence on the ansatz, is left for future work.

\subsubsection{Future work}

More broadly, the OSE perspective raises several open questions.
We have demonstrated above using OSE regularization to probe the tradeoff between energy accuracy and operator non-stabilizerness.
However, this should be distinguished from the more ambitious task of designing low-T-gate state-preparation algorithms. Converting this tradeoff into explicit circuit constructions with reduced T-gate counts remains an open problem.

Another important question is the scaling of OSE in optimized ground-state-preparation circuits.
While an extensive number of recent works have studied the non-stabilizerness of the states in~\cite{tarabunga2023many,odavic2023complexity,falcao2025nonstabilizerness,hoshino2026stabilizer,timsina2025robustness,
keeble2026neural} and out of equilibrium~\cite{turkeshi2025magic,turkeshi2025pauli,tirrito2024anticoncentration,falcao2025nonstabilizernessdynamics,aditya2025mpemba,aditya2025growth},
the non-stabilizerness of operators is an under-explored area of study.
For example, it remains to be understood whether the OSE remains bounded in gapped phases, grows near criticality, or depends sensitively on the freedom in the choice of state-preparation unitary discussed above.
We have shown that one crucial issue needs to be addressed for such a study is the extra degree of freedom leading to widespread OSE values. 
Finally, the truncation procedures used in SPD simulation, such as coefficient-threshold or operator-weight truncation~\cite{li2026dual,lerch2026efficient}, can be viewed as forms of implicit regularization.
Clarifying their relation to explicit OSE regularization may lead to more principled optimization strategies that balance energy accuracy, classical simulability, and non-stabilizerness.

\subsection{\label{subsec: parallelization} Parallelization}

Efficient parallelization is crucial for making SPD optimization practical, especially in variational settings where cost functions and gradients must be evaluated repeatedly.
Even a modest order-of-magnitude speedup can shift an optimization from weeks to days.
The parallelizability of Pauli propagation is sometimes obscured by the fact that the dominant operations involve data-structure tasks such as searching, sorting, and merging Pauli strings.
Some previous works have either not focused on parallel implementation~\cite{loizeau2025quantum} or have emphasized the challenges associated with parallelizing these operations~\cite{rudolph2025pauli}.
A related statement is that Pauli-propagation simulations often become memory-limited before becoming compute-limited: even single-threaded implementations can saturate the available memory~\cite{rudolph2025pauli}.

While memory usage and Pauli-string merging are indeed the main limiting factors for SPD simulation, we would like to emphasize that these bottlenecks do not preclude useful parallel speedup.
In fact, prior works have already demonstrated speedups with shared-memory parallelization on a single node~\cite{beguvsic2024fast,beguvsic2025real}.
More recently, Ref.~\cite{broers2025scalable} has further demonstrated efficient large-scale parallelization by distributing Pauli strings across multiple nodes using hash maps, exhibiting both strong and weak scaling up to $2^{17}$ parallel processes.

In this work, we implement the SPD simulation using the JAX library and observe a noticeable speedup on a single GPU relative to the corresponding CPU execution.
Both forward and backward propagation are equally parallelizable because they share the same primitive operations: applying Pauli rotations, generating conjugated strings, sorting or searching, and merging coefficients.
We therefore observe similar performance for forward and backward propagation.
The benchmark setup and results are discussed in Appendix~\ref{appendix: implementation}.
Recently, NVIDIA has also introduced Pauli-propagation functionality in the cuQuantum library~\cite{bayraktar2023cuquantum}.
In general, multi-GPU parallelization is also possible following the setup in Ref.~\cite{broers2025scalable}.

\section{\label{sec: conclusion} Conclusion}

In this work, we developed a backpropagation algorithm for evaluating gradients of cost functions with respect to circuit parameters within the framework of SPD simulation.
The algorithm follows the idea of reverse-mode automatic differentiation, but exploits the reversibility of quantum circuits to approximately reconstruct intermediate states on the fly during the backward pass, instead of storing them.
We find that the finite truncation errors accumulated in the forward and backward passes are of the same order of magnitude, yielding an approximate but controlled gradient estimate.
The resulting algorithm reduces the memory cost from $\mathcal{O}(N_P \ngate)$ for standard reverse-mode automatic differentiation to $\mathcal{O}(N_P)$, where $N_P$ is the size of the largest SPO and $\ngate$ is the number of parameters.
This reduction is particularly important because memory is typically the main bottleneck in SPD simulation, and it does not come at the expense of asymptotic runtime, since derivative backpropagation is required in both approaches. Compared to finite-difference gradient methods, our method has a $\mathcal{O}(n_{\rm param})$ speedup because it does not need to repeat a SPD simulation for each of the $n_{\rm param}$ parameters.
The framework is generic and supports a range of applications in the classical optimization of quantum circuits.

We present two prominent applications: low-energy state preparation and time-evolution circuit compression.
For the first application, we validate our approach by comparing it with the state-of-the-art tensor network approaches, including matrix product states (MPS) and projected entangled-pair states (PEPS) methods, for preparing ground states of 1D and 2D transverse-field Ising (TFI) models. We find that our proposed method achieves comparable accuracy to the recent MPS~\cite{jobst2022finite,sokolov2025bang} and PEPS~\cite{zhang2024bang} based results. 
To demonstrate the flexibility of the SPD-based approach, we apply the same algorithm to find state-preparation circuits for TFI and Heisenberg models on 3D lattices.
Finally, we show that time-evolution circuit compression fits naturally within this framework by optimizing compressed circuits for 2D TFI dynamics.

Our results position SPD with backpropagation as a complementary approach to tensor-network-based classical optimization methods for quantum circuits~\cite{lin2021real,jobst2022finite,zhang2024bang,sokolov2025bang,gibbs2025deep,watanabe2026tensor}.
Compared with tensor network methods, SPD offers several distinct advantages.
First, its computational complexity is determined primarily by the operator non-stabilizerness rather than by state entanglement, making it promising for optimizing circuits with large state entanglement but low operator non-stabilizerness.
Second, the SPD implementation is agnostic to circuit layout and system geometry, allowing the same implementation to be applied to circuits on arbitrary graphs. This makes it especially useful for exploring higher-dimensional models on irregular geometries.
Third, SPD can naturally handle rotations generated by high-weight operators and long-range gates, suggesting potential applications to ans\"atze whose classical optimization is challenging for tensor-network methods.
These features suggest several future directions. 
For example, high-weight circuit optimization could be useful for lattice gauge theory simulations in higher dimensions, where the relevant operators can have high weight.
Another promising direction is the Trotterized multiscale entanglement renormalization ansatz~\cite{miao2025convergence} for state preparation, whose long-range gates have so far limited classical optimization studies beyond one dimension.

More broadly, SPD simulation provides a natural framework for studying quantum complexity in the Heisenberg picture.
In this setting, the operator stabilizer R\'enyi entropy is both a complexity measure for sparse Pauli simulation and a magic monotone for Heisenberg-evolved operators.
Our OSE-regularization results provide a first demonstration of how SPD optimization can be used to explore the tradeoff between energy accuracy and operator non-stabilizerness.
This does not yet constitute an algorithm for low-T-gate state preparation, since translating reduced OSE into explicit reductions in non-Clifford gate counts remains open.
Nevertheless, the framework developed here provides a concrete starting point for optimization methods that account not only for variational accuracy, but also for classical simulability and non-Clifford resource requirements.

{\bf Data Availability:} The data of this work will be provided on Zenodo at publication.

{\bf Code Availability:} The Python code with JAX and NumPy implementation is available on GitHub at Ref.~\cite{github2026}.

\begin{acknowledgments}

S.L. thanks Lode Pollet for the instruction on the ALPS package~\cite{bauer2011alps} setup for the QMC calculation. S.L. thanks Bernhard Jobst and Cheng-Ju Lin for providing the numerical data for comparison. S.L. thanks Poetri Sonya Tarabunga, Mohsin Iqbal, and Cristina Cirstoiu for helpful input and discussion. 

\end{acknowledgments}

\appendix

\section{Finite-depth circuit energy data}
\label{appendix: energy}

Here we report the energy data obtained from our SPD optimization over finite-depth circuits.
We consider the following two different setups. 
One targets the thermodynamic limit of infinite system size, while the other targets a given finite size.

In the thermodynamic-limit setup, we evaluate the energy density $E = \langle H \rangle /N$ of the Hamiltonian $H$ on a PBC lattice whose linear size $L$ is larger than the circuit light cone. We choose $L = 2 \times \text{layer} +2$.
Due to the translational invariance of both the Hamiltonian and the circuit and the finite depth, the evaluated energy density is the same for any larger $L$.
Hence, we compare the results with increasing circuit depth to the value in the thermodynamic limit.

\subsubsection*{1D TFI}
We begin with the 1D TFI model at $g=1.1$.
Table~\ref{tab:TN_vs_SPD_inf1d} compares the finite-depth HVA energies obtained with SPD backpropagation against the MPS-based optimization of Ref.~\cite{sokolov2025bang}; the lattice is larger than the circuit light cone, so these are thermodynamic-limit energy densities.

\begin{table}[htbp]
\centering
\begin{tabular*}{\columnwidth}{@{\extracolsep{\fill}} c cc }
 \hline
 & MPS optimization~\cite{sokolov2025bang} & SPD optimization \\
 layer & $E_\text{circ}$ & $E_\text{circ}$ \\ [0.25em]
 \hline\hline
 2 & -1.324301 & -1.3243022 \\
 3 &-1.334014 & -1.3340151 \\
 4 & -1.338055& -1.3380548 \\
 5 & -1.340026& -1.3400265 \\
 6 & -1.341092& -1.3410928 \\
 7 & -1.341711& -1.3417112 \\
 8 & -1.342090& -1.3420908 \\
 9 & -1.342332& -1.3423321 \\
 10 &-1.342491& -1.3424913 \\
 \hline\hline
\multicolumn{3}{c}{$E_\text{exact}=-1.342864$} \\
 \hline
\end{tabular*}
\caption{
Energy density of 1D Transverse field Ising model with $g=1.1$.
The two column groups report finite-depth circuit energies obtained by optimizing the same HVA with MPS- and SPD-based gradients, respectively.
The final row is the exact ground-state energy density in the thermodynamic limit, not a circuit-optimization result.
}
\label{tab:TN_vs_SPD_inf1d}
\end{table}

\subsubsection*{2D TFI}
We next consider the 2D TFI model at $g=3.1$.
Table~\ref{tab:TN_vs_SPD_inf2d} separates two optimizations of the same HVA: one using iPEPS-based gradients and one using SPD backpropagation.
The iPEPS ground-state energy is included only as a thermodynamic-limit reference.

\begin{table}[htbp]
\centering
\begin{tabular*}{\columnwidth}{@{\extracolsep{\fill}} c cc cc }
  \hline
 & \multicolumn{2}{c}{iPEPS optimization~\cite{zhang2024bang}} & \multicolumn{2}{c}{SPD optimization} \\
 layer & $E_\text{circ}$ & $\epsilon_\text{TN}$ & $E_\text{circ}$ & $\epsilon_\text{SPD}$ \\ [0.25em]
 \hline\hline
 2 & -3.274164 & 0 & -3.274167 & $2.9\times 10^{-5}$ \\
 3 & -3.278906 & 0  & -3.278821 & $9.5\times 10^{-4}$ \\
 4 &  -3.28142 & $1.2\times 10^{-6}$ & -3.281067 & $4.6 \times 10^{-3}$\\
 5 &  -3.28170 & $3.2\times 10^{-6}$ & & \\
 6 &  -3.28182 & $3.8\times 10^{-6}$ & & \\
 \hline\hline
\multicolumn{5}{c}{$E_\text{iPEPS}=-3.2844772$} \\
 \hline
\end{tabular*}
\caption{
Energy density of 2D Transverse field Ising model with $g=3.1$.
The two column groups report finite-depth circuit energies obtained by optimizing the same HVA with iPEPS- and SPD-based gradients, respectively.
$\epsilon_\text{TN}$ is the numerical uncertainty reported in Ref.~\cite{zhang2024bang}, whereas $\epsilon_\text{SPD}$ is the empirical SPD truncation-error estimate defined and tested in Appendix~\ref{appendix: error}.
The final row is a ground-state reference, not a circuit-optimization result.
}
\label{tab:TN_vs_SPD_inf2d}
\end{table}

\subsubsection*{3D TFI}
For the 3D TFI model at $g=5.2$, we instead target a fixed $8\times8\times8$ PBC lattice.
Table~\ref{tab: SPD_TFI_3D} reports the SPD-optimized circuit energies and compares them with a QMC calculation on the same finite lattice.

\begin{table}[htbp]
\centering
\begin{tabular*}{\columnwidth}{@{\extracolsep{\fill}} c c c c }
 \hline
 method & layer & energy density & error estimate \\
 \hline\hline
 SPD & 1 & -5.3392745 & $5.2797 \times 10^{-6}$  \\
 SPD & 2 &-5.3531064 & 0.0002 \\
 SPD & 3 & -5.3585820 &  0.3791 \\
 \hline\hline
QMC reference & --- &  -5.3587297 & $7.9312 \times 10^{-5}$ \\
 \hline
\end{tabular*}
\caption{
Energy density of the 3D Transverse field Ising model with $g=5.2$ on an $8\times8\times8$ PBC lattice.
For the optimized circuits, the last column is the empirical SPD truncation-error estimate defined in Appendix~\ref{appendix: error}; for QMC, it is the statistical uncertainty.
The QMC entry is a benchmark and was not obtained by circuit optimization.
}
\label{tab: SPD_TFI_3D}
\end{table}

\subsubsection*{3D AFH}
Finally, for the 3D AFH model, we optimize a circuit on a $4\times4\times4$ PBC lattice and then reevaluate the same parameters on lattices larger than the circuit light cone.
Table~\ref{tab: SPD_AFH_3D} therefore reports both the finite-size optimization result and the effectively thermodynamic-limit circuit energy, together with the corresponding QMC references.

\begin{table}[htbp]
\centering
\begin{tabular*}{\columnwidth}{@{\extracolsep{\fill}} c c }
 \hline
 setup & energy density \\ 
 \hline \hline
 $4\times4\times4$ PBC optimization & -0.889315 \\
 $L\geq 8$ light-cone evaluation & -0.8891 \\
 \hline \hline
 QMC ($4\times 4\times 4$)  & -0.91588(75) \\
QMC thermodynamic limit~\cite{vlaar2021simulation} & -0.902325(11) \\
 \hline
\end{tabular*}
\caption{
Energy density of the 3D anti-ferromagnetic Heisenberg model.
The same optimized circuit parameters are evaluated on systems larger than the circuit light cone to obtain the effectively thermodynamic-limit SPD estimate.
We compare our SPD result to both the finite size QMC simulation and the extrapolated value in the literature~\cite{vlaar2021simulation}.
}
\label{tab: SPD_AFH_3D}
\end{table}

\twocolumngrid
\section{Empirical error estimate}
\label{appendix: error}

In practice, SPD simulation is performed with finite truncation error.
A rigorous analysis of observable-error bounds for Pauli propagation is given in Ref.~\cite{rudolph2025pauli}.
Here, we only summarize the distinction between those worst-case bounds and the empirical error estimate used in this work.

We use the rescaled Hilbert-Schmidt inner product
\begin{equation}
    \langle A, B \rangle_{\mathrm{rHS}} := 2^{-n}\mathrm{Tr}[A^\dagger B],
\end{equation}
under which Pauli strings form an orthonormal basis.
Thus the coefficients in the decomposition
\(
O = \sum_i a_i P_i
\)
inherit the $\ell^2$ norm.
For an expectation value in the computational-basis state, we can write
\(
\text{Tr}[|0\rangle\langle 0 | U^\dagger O U ] = \langle w | v(t) \rangle
\)
in the vectorized operator space, where $|v(t)\rangle$ is the coefficient vector of the time-evolved operator and $|w\rangle = 2^n |0\rangle\langle 0 |$ absorbs the rHS normalization.
With this convention,
\(
\langle w | w \rangle = \text{Tr}[(2^n|0\rangle \langle 0 |)^2  / 2^{n} ] = 2^n
\).

For each gate application, let the exact and truncated evolutions be
\begin{equation}
    |v_s\rangle = U_s |v_{s-1}\rangle, \qquad
    |\tilde v_s\rangle = P_s U_s |\tilde v_{s-1}\rangle .
\end{equation}
Here $U_s$ is the unitary coefficient-space update induced by the gate, and $P_s$ is the projector onto the Pauli strings retained by the truncation rule at step $s$.
The discarded $\ell^2$ weight at this step is
\begin{equation}
    \epsilon^\text{trunc}_{s}
    :=
    \|(\mathbbm{1} - P_s) U_s |\tilde v_{s-1}\rangle\|
    =
    \sqrt{\sum_{P_i\in \mathcal{D}_s} |a^{[s]}_{P_i}|^2},
    \label{eq: local_truncation_error}
\end{equation}
where $\mathcal{D}_s$ is the set of Pauli strings discarded at step $s$.

Define the error
\begin{equation}
    \epsilon_s := \|\,|v_s\rangle - |\tilde v_s\rangle\|.
\end{equation}
Then
\begin{equation}
    |v_s\rangle - |\tilde v_s\rangle
    = U_s (|v_{s-1}\rangle - |\tilde v_{s-1}\rangle)
    + (\mathbbm{1} - P_s) U_s |\tilde v_{s-1}\rangle,
\end{equation}
By the triangle inequality, we have
\begin{equation}
  \Big \lVert  |{v}_s\rangle -    |\tilde{v}_s \rangle \Big\rVert \leq
  \Big\lVert |v_{s-1}\rangle - |\tilde{v}_{s-1}\rangle \Big\rVert + \epsilon_s^\text{trunc}
\end{equation}
That is
\begin{equation}
    \epsilon_s \leq \epsilon_{s-1} + \epsilon^\text{trunc}_{s}
\end{equation}
Thus,
\begin{equation}
    \epsilon_t \le \sum_{s=1}^t \epsilon_s^{\mathrm{trunc}} .
    \label{eq: l2_worst_case}
\end{equation}
The corresponding observable-error bound follows from Cauchy--Schwarz,
\begin{equation}
    |\langle w|v(t)\rangle - \langle w|\tilde v(t)\rangle|
    \le \|w\| \, \epsilon_t.
\end{equation}
For $w = 2^n |0\rangle\langle 0|$, one has
\begin{equation}
    \|w\|^2 = 2^{-n}\mathrm{Tr}(w^\dagger w) = 2^n,
    \quad \Rightarrow \quad \|w\| = 2^{n/2}.
\end{equation}
Hence
\begin{equation}
    |\langle w|v(t)\rangle - \langle w|\tilde v(t)\rangle|
    \le 2^{n/2} \sum_{s} \epsilon_s^{\mathrm{trunc}}.
    \label{eq: l2_observable_worst_case}
\end{equation}
This is a valid worst-case bound, but it is typically far too loose to be useful as a practical error bar.

A tighter rigorous estimate can be obtained from the $\ell^1$ norm of the discarded Pauli coefficients~\cite{rudolph2025pauli}. 
Let us denote the truncation residual at step $s$ by
\(
\Delta O_s = \sum_{P_i \in \mathcal{D}_s} a^{[s]}_{P_i} P_i.
\)
Then for any density matrix $\rho$,
\begin{equation}
    \mathrm{Tr}(\Delta O_s \rho)
    = \sum_{P_i \in \mathcal{D}_s} a^{[s]}_{P_i} \mathrm{Tr}(P_i \rho),
\end{equation}
and using
\begin{equation}
    |\mathrm{Tr}(P_i \rho)| \le 1,   \quad \forall P_i
\end{equation}
we obtain
\begin{equation}
    \big|\mathrm{Tr}(\Delta O_s \rho)\big|
    \le \sum_{P_i \in \mathcal{D}_s} |a^{[s]}_{P_i}|.
\end{equation}
Accumulating this quantity over truncation steps gives the worst-case estimate
\begin{equation}
    \big|\mathrm{Tr}(O \rho) - \mathrm{Tr}(\tilde O \rho)\big|
    \le \sum_s \sum_{P_i \in \mathcal{D}_s} |a^{[s]}_{P_i}|.
\end{equation}
In our numerical experiments, however, even this bound remains conservative compared with the observed errors.

The empirical observable-error estimate used in this work is not the worst-case bound in Eq.~\eqref{eq: l2_observable_worst_case}.
In particular, we do not include the factor $2^{n/2}$ from the Cauchy--Schwarz step.
Furthermore, we directly use the root-sum-square of the discarded $\ell^2$ weights,
\begin{equation}
    \epsilon_\text{emp}
    :=
    \left(
    \sum_{s=1}^t
    \left(\epsilon_s^{\mathrm{trunc}}\right)^2
    \right)^{1/2}.
    \label{eq: empirical_truncation_error}
\end{equation}
as the practical estimate for the observable error.
There are two empirical ingredients behind this choice.
First, truncation residuals created at different steps are not generally aligned after subsequent propagation, so their accumulated effect is better described by quadrature than by the coherent sum in Eq.~\eqref{eq: l2_worst_case}.
Second, the accumulated truncation error is not observed to be worst-case aligned with the measurement vector $w$; this is why the factor $2^{n/2}$ in Eq.~\eqref{eq: l2_observable_worst_case} is not included in the practical estimate.
These assumptions are not guaranteed, and Eq.~\eqref{eq: empirical_truncation_error} should therefore be interpreted as an empirical observable-error estimate, not as a rigorous bound.

\begin{figure}
    \centering
    \includegraphics[width=\linewidth]{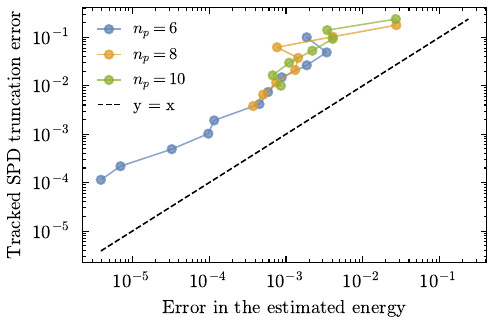}
    \caption{
    Numerical test of the empirical SPD error estimate for fixed 2D TFI HVA circuits with three, four, and five layers ($n_\text{param}=6,8,10$).
    The circuit parameters are taken from the iPEPS optimization of Ref.~\cite{zhang2024bang}; only the SPD truncation threshold is varied along each curve.
    The horizontal axis is the observed error $|E_\text{SPD}-E_\text{iPEPS}|$, treating the iPEPS evaluation of the same circuit as exact, and the vertical axis is the tracked estimate obtained by accumulating discarded $\ell^2$ weights in quadrature according to Eq.~\eqref{eq: empirical_truncation_error}.
    The dashed line denotes equality.
    }
    \label{fig: truncation error}
\end{figure}

To test this estimate numerically, we take the optimized HVA parameters reported by the iPEPS study of Ref.~\cite{zhang2024bang} and hold them fixed.
We consider circuits with three, four, and five layers, corresponding to $n_\text{param}=6$, $8$, and $10$ variational parameters, respectively.
For each circuit, we reevaluate the energy with SPD over a sequence of truncation thresholds and record both the empirical estimate $\epsilon_\text{emp}$ and the resulting SPD energy.
We treat the iPEPS evaluation of that same fixed circuit as the reference value and define the observed error as $|E_\text{SPD}-E_\text{iPEPS}|$.
As shown in Fig.~\ref{fig: truncation error}, for the examples considered here, we observed the empirical error estimates upper bounds the error and hence provides a useful practical diagnostic for SPD expectation values.
This is the error estimate reported throughout the main text.

\section{Implementation}
\label{appendix: implementation}

In this appendix, we describe the implementation details relevant for the parallelization discussion in Sec.~\ref{subsec: parallelization}.
The main point is that the formulation used to compute gradients, based on conjugated Pauli pairs $(P,Q)$, is compatible with parallel execution.
It is not necessary to abandon this pair structure in order to obtain a GPU-friendly implementation.

Our JAX implementation stores the SPO as a lexicographically sorted array of packed $X/Z$ bit strings, together with the corresponding coefficient array.
For a Pauli rotation generated by $\sigma$, the update proceeds as follows.
First, the code computes which stored Pauli strings anticommute with $\sigma$.
For every stored string $P$, it then computes the conjugated string $Q\propto \sigma P$, up to the phase convention used in Sec.~\ref{subsec: spd}, and searches for $Q$ in the sorted array.
If both members of a pair are already present, their coefficients are updated by the $2\times2$ rotation in Eq.~\eqref{eq: forward}.
If the conjugated partner is absent, a new row is inserted with the corresponding coefficient.
Finally, coefficients below the threshold are discarded and the array is sorted again.
This is the search-update-merge algorithm used for the data shown below.

The backward pass uses the same data-movement pattern.
The only difference is that the sparse operator stores both the coefficients and the coefficient derivatives.
After the same partner search, the code updates both arrays and accumulates the parameter derivative from the cross terms in Eq.~\eqref{eq: gradient}.
Thus the gradient calculation adds local arithmetic but does not introduce a different global communication pattern.
Forward and backward propagation are therefore parallelizable in the same way.
The crucial implementation point is that the $(P,Q)$ pair structure needed for the gradient is also the structure used by the parallel update.

We benchmark the implementation on a two-dimensional global-quench problem following the setup of Ref.~\cite{beguvsic2024fast}.
The Hamiltonian is
\begin{equation*}
    H = -\sum_{\langle i,j\rangle} Z_iZ_j - h\sum_i X_i
\end{equation*}
on an $11\times 11$ open-boundary square lattice, with $h=3.044382$.
The initial observable is a single $X$ operator at the center site.
We use a first-order product formula with time step $dt=0.04$ and coefficient threshold $\delta=2^{-18}$.
The comparison data are generated with the same first-order setup, and the throughput plot uses the completed time steps before the run is stopped, for example because of memory limitations.
After each time step, we record the number of retained Pauli strings and define the average throughput as
\begin{equation}
    \frac{\bar N_P\, n_\text{gate}}{\Delta t_\text{wall time}},
\end{equation}
where $\bar N_P$ is the average of the number of strings before and after the step, $n_\text{gate}=2L(L-1)+L^2$ is the number of Pauli rotations in one time step, and $\Delta t_\text{wall time}$ is the measured wall time for that step.

\begin{figure}[ht]
\includegraphics[width=0.5\textwidth]{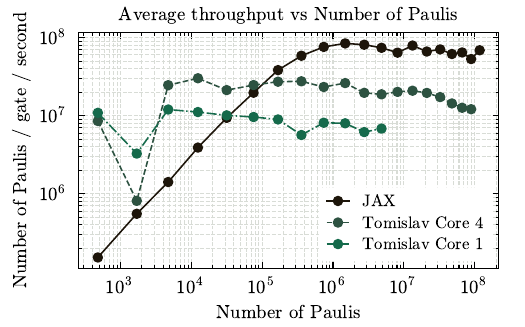}
\caption{
Average throughput for the two-dimensional global-quench benchmark as a function of the number of retained Pauli strings.
The JAX implementation uses the search-update-merge algorithm described in the text, with the data generated on a single NVIDIA H200 GPU.
The CPU reference data are generated with the implementation of Ref.~\cite{beguvsic2024fast} using one and four CPU cores.
The comparison shows that the $(P,Q)$ pair formulation used for gradient extraction remains compatible with parallel execution.
}
\label{fig: speed}
\end{figure}

The result is shown in Fig.~\ref{fig: speed}.
The benchmark is not intended as an exhaustive performance study.
Rather, it demonstrates that the implementation based on conjugated Pauli pairs can be parallelized and reaches comparable throughput to an existing CPU implementation on the same benchmark task.
Furthermore, we have also run this benchmark with our own CPU implementation written in Rust using hash-table based distributed data structures in the  spirit of Ref.~\cite{broers2025scalable} (not shown on the figure). 
Interestingly, although the two CPU implementations rely on different data structures and parallelization procedures, we found surprisingly similar performance behavior, including a slowdown using four cores when the number of Pauli strings becomes high. We could also confirm that the gradient calculation benefited as much as the forward propagation from the parallelization strategy based on hash tables. 
We suspect that further improvements are possible, for example by reducing the communication between cores when using CPUs, or by adapting the ideas of Ref.~\cite{broers2025scalable} to GPU implementations.

\onecolumngrid
\section{Numerical Precision}
\label{appendix: numerical_precision}

In this appendix, we examine the effect of floating-point precision on the SPD optimization.
This effect is visible in the 1D TFI benchmark of Sec.~\ref{subsec: result_state_prep}, especially for deeper HVA circuits.
The HVA circuit is a fermionic Gaussian circuit, so the corresponding SPD simulation is exact without coefficient truncation.
Consequently, this isolates numerical precision effects from truncation errors.

We focus on the 10-layer HVA ansatz and use 10 random initializations.
For each initialization, we run the optimization with two backend implementations (JAX and NumPy), two floating-point precisions (FP32 and FP64), and two Hamiltonian representations (a local Hamiltonian term and the full Hamiltonian).
This gives eight optimization settings for each initialization.
All runs use the SciPy L-BFGS-B optimizer with default parameters.
After convergence, we restart the optimization from the obtained parameters with tighter tolerances, $\mathrm{gtol}=10^{-7}$ and $\mathrm{ftol}=10^{-9}$, to reduce the chance that the observed differences come from premature convergence or from the approximate inverse Hessian history.

\begin{figure}[ht]
\includegraphics[width=1.\textwidth]{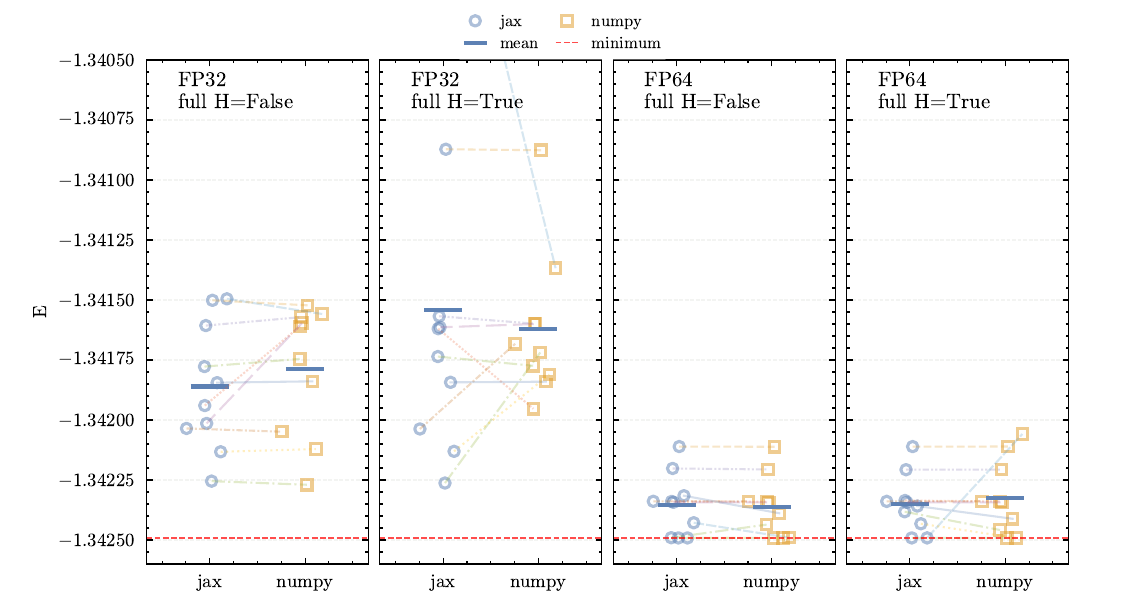}
\caption{
Energy-density optimization results for the 1D TFI model using a 10-layer HVA circuit.
For 10 fixed random initializations, we compare different optimization settings: (i) single (FP32) versus double precision (FP64), (ii) JAX versus NumPy implementation, and (iii) optimization using either a local Hamiltonian term or the full Hamiltonian.
Small horizontal jitter is added to distinguish otherwise overlapping points.
Optimizations starting from the same random initialization are connected by light lines between the JAX and NumPy implementations.
The blue markers indicate the mean, and the red dashed line indicates the expected minimum value.
}
\label{fig: single_versus_double}
\end{figure}

The results are shown in Fig.~\ref{fig: single_versus_double}.
The dominant factor controlling the final energy is the floating-point precision.
In contrast, the differences between JAX and NumPy, and between the local-term and full-Hamiltonian formulations, are comparatively small.

We think the reason single precision optimization fails is due to precision loss in the gradient calculation.
Each gradient component is obtained as a sum of contributions from many cross terms of coefficients and derivatives of Pauli strings, and cancellations between these contributions can amplify roundoff error.
The residual differences between backend implementations and Hamiltonian representations can be understood in the same way: different operation graphs accumulate floating-point error in slightly different ways.

\newpage

\twocolumngrid
\section{Local $X,Z$ error}
\label{sec: local cost}

In this appendix, we provide a simple derivation of the lower and upper bound between the local $X,Z$ cost used in the main text and the Hilbert-Schmidt test cost.
The derivation is deterministic and uses only the Pauli-basis expansion of the relative unitary.
It is therefore simpler than the route based on locally scrambling ensembles and risk equivalences used in Refs.~\cite{caro2023out,d2025circuit}.
Using only $X$ and $Z$ generators is sufficient because every non-identity single-qubit Pauli anticommutes with at least one of them.

Let $V$ denote the target unitary, let $U(\boldsymbol{\theta})$ denote the variational unitary, and let $d=2^n$.
The Hilbert-Schmidt test cost is
\begin{equation}
    \mathcal{L}_\text{HST}
    =
    1 - \frac{1}{d^2}
    \left \lvert \mathrm{Tr}\!\left[U(\boldsymbol{\theta})^\dagger V\right] \right \rvert^2 .
\end{equation}
The local cost considered in the main text is
\begin{align}
\mathcal{L}_{X,Z}
&=
    \sum_{i=1}^n \left\lVert X_i(t) - \tilde{X}_i(t) \right\rVert_\text{rHS}^2
    +
    \sum_{i=1}^n \left\lVert Z_i(t) - \tilde{Z}_i(t) \right\rVert_\text{rHS}^2 ,
\end{align}
where
\begin{align}
    X_i(t) &= U(\boldsymbol{\theta})^\dagger X_i U(\boldsymbol{\theta}),&
    Z_i(t) &= U(\boldsymbol{\theta})^\dagger Z_i U(\boldsymbol{\theta}),\\
    \tilde{X}_i(t) &= V^\dagger X_i V,&
    \tilde{Z}_i(t) &= V^\dagger Z_i V .
\end{align}
Define the relative unitary
\begin{equation}
    W = V U(\boldsymbol{\theta})^\dagger .
\end{equation}
For any single-site generator $G\in\{X_i,Z_i\}_{i=1}^n$, unitary invariance of the rescaled Hilbert-Schmidt norm gives
\begin{align}
    \left\lVert U(\boldsymbol{\theta})^\dagger G U(\boldsymbol{\theta})
    -
    V^\dagger G V
    \right\rVert_\text{rHS}
    &=
    \left\lVert W^\dagger G W - G \right\rVert_\text{rHS} \nonumber\\
    &=
    \left\lVert [G,W] \right\rVert_\text{rHS}.
\end{align}
We now expand $W$ in the Pauli basis,
\begin{equation}
    W = \sum_{Q\in\mathcal{P}_n} c_Q Q,
    \qquad
    c_Q = 2^{-n}\mathrm{Tr}(QW).
\end{equation}
Since $W$ is unitary and Pauli strings are orthonormal in the rescaled Hilbert-Schmidt inner product,
\begin{equation}
    \sum_Q |c_Q|^2 = 1.
\end{equation}
The identity coefficient satisfies
\begin{equation}
    |c_{\mathbbm{1}}|^2
    =
    \frac{1}{d^2}\left|\mathrm{Tr}(W)\right|^2
    =
    \frac{1}{d^2}
    \left|\mathrm{Tr}\!\left[U(\boldsymbol{\theta})^\dagger V\right]\right|^2,
\end{equation}
and therefore
\begin{equation}
    \mathcal{L}_\text{HST}
    =
    \sum_{Q\neq \mathbbm{1}} |c_Q|^2 .
\end{equation}

For a fixed generator $G$, only Pauli strings that anticommute with $G$ contribute to the commutator:
\begin{equation}
    \left\lVert [G,W]\right\rVert_\text{rHS}^2
    =
    4\sum_{Q:\{Q,G\}=0} |c_Q|^2 .
\end{equation}
Summing this expression over $G\in\{X_i,Z_i\}_{i=1}^n$ gives
\begin{equation}
    \mathcal{L}_{X,Z}
    =
    4\sum_{Q\in\mathcal{P}_n} N_{X,Z}(Q)\, |c_Q|^2 ,
\end{equation}
where $N_{X,Z}(Q)$ is the number of single-site generators in $\{X_i,Z_i\}_{i=1}^n$ that anticommute with $Q$.
For the identity string, $N_{X,Z}(\mathbbm{1})=0$.
For every non-identity Pauli string,
\begin{equation}
    1 \leq N_{X,Z}(Q) \leq 2n,
    \qquad Q\neq\mathbbm{1}.
\end{equation}
The lower inequality holds because any non-identity single-site factor anticommutes with at least one of $X_i$ or $Z_i$, while the upper inequality is immediate from the number of generators.
It follows that
\begin{equation}
    4\mathcal{L}_\text{HST}
    \leq
    \mathcal{L}_{X,Z}
    \leq
    8n\,\mathcal{L}_\text{HST}.
\end{equation}
Equivalently,
\begin{equation}
    \frac{1}{8n}\mathcal{L}_{X,Z}
    \leq
    \mathcal{L}_\text{HST}
    \leq
    \frac{1}{4}\mathcal{L}_{X,Z}
\end{equation}
for the summed $X,Z$ cost used in this work.
If the local cost is instead averaged over the $2n$ generators, the explicit factor of $n$ in the lower bound is absorbed into the definition of the averaged cost.

\section{Gradient calculation through forward propagation}
\label{appendix: forward_mode}

In this section, we follow the idea of forward-mode automatic differentiation and explain the resulting formalism when applied to SPD simulation.
We show that one can obtain the gradient through a single ``forward pass''. This method avoids the need for the backpropagation step at the cost of significantly increasing the memory footprint by a factor equal to the number of parameters $\nparam$.

The overall idea is to propagate the operator coefficients forward as usual, while simultaneously propagating all the partial derivatives of the coefficients with respect to the previous gate angles
\[
\frac{\partial a_{P_m}^{[t]}}{\partial \theta_j}, \qquad \forall j<t
\]
forward and finally obtain the gradient at the last step by utilizing the chain rule.

Define 
\[
s_{{P_m};j}^{[t]}
:=
\frac{\partial a_{P_m}^{[t]}}{\partial \theta_j}.
\]
Differentiate the pair update in Sec.~\ref{subsec: spd} with respect to $\theta_j$:
\[
\begin{pmatrix}
a_{P_m}^{[t]} \\
a_{Q_m}^{[t]}
\end{pmatrix}
=
 \mathcal{U}(\theta_{t})
\begin{pmatrix}
a_{P_m}^{[t-1]} \\
a_{Q_m}^{[t-1]}
\end{pmatrix}
\]

We have for $j\leq t-1$ 
\[
\begin{pmatrix}
s_{{P_m};j}^{[t]} \\
s_{{Q_m};j}^{[t]}
\end{pmatrix}
= \mathcal{U}(\theta_{t})
\begin{pmatrix}
s_{{P_m};j}^{[t-1]} \\
s_{{Q_m};j}^{[t-1]}
\end{pmatrix}
\]
and for $j=t$ 
\[
\begin{pmatrix}
s_{{P_m};j}^{[t]} \\
s_{{Q_m};j}^{[t]}
\end{pmatrix}
= 
\frac{    \partial \mathcal{U}(\theta_{t})
}{\partial \theta_j}
\begin{pmatrix}
a_{P_m}^{[t-1]} \\
a_{Q_m}^{[t-1]}
\end{pmatrix}.
\]
The two formulas above provide the initialization of the partial derivatives $s_{{P_m};j}^{[t]}$ at $j=t$ and the update rule on how to propagate all previous partial derivatives $s_{{P_m};j}^{[t-1]}$ for $j\leq t-1$ forward from $t-1$ to $t$.
Since the update rule of the partial derivatives $s_{{P_m};j}^{[t-1]}$ is exactly the same as the coefficient update rule, one can store and update them together with the coefficients. 
The drawback is the growing size of the SPO with the additional partial derivative terms attached.
This iteration is carried out to the final layer.

For a cost function depending on the final coefficients:
\[
\mathcal L(\theta)
= f( \{ a_{P_m}^{[T]} \})
\]

The gradient with respect to the parameter $\theta_j$ is given by the chain rule:
\[
\frac{\partial \mathcal L}{\partial \theta_j}
=
\sum_m  \frac{\partial f( \{ a_{P_m}^{[T]} \})} {\partial a_{P_m}^{[T]} }
\frac{\partial a_{P_m}^{[T]}}{\partial \theta_j}.
\]

This method of gradient calculation imposes to keep in memory all the partial derivatives of the coefficients during the forward propagation, effectively multiplying the memory cost by a factor equal to the number of variational parameters $\nparam$.

\bibliography{ref}%

\end{document}